\documentclass[useAMS,usenatbib]{mn2e}

\usepackage{booktabs,multicol}
\usepackage{color}
\usepackage{multirow}
\usepackage{braket}
\usepackage{graphicx}
\usepackage{cite}
\usepackage{bm}
\usepackage[caption=false]{subfig}

\title[Important processes involving Cobalt ions]{Photoionization of Co$^+$ and electron-impact excitation of Co$^{2+}$ using the Dirac $R$-matrix method}
\author[N. B. Tyndall, C. A. Ramsbottom, C. P. Ballance and A. Hibbert]{N. B. Tyndall$^{1}$\thanks{E-mail:
ntyndall01@qub.ac.uk}, C. A. Ramsbottom$^{1}$, C. P. Ballance$^{1}$ and A. Hibbert$^{1}$\\
$^{1}$School of Mathematics \& Physics,\\
The Queen's University of Belfast, Belfast,\\
BT7 1NN,\\
Northern Ireland\\}
\begin{document}

\date{Accepted (date). Received (date); in original form (date)}

\pagerange{\pageref{firstpage}--\pageref{lastpage}} \pubyear{2002}

\maketitle

\label{firstpage}


\begin{abstract}
Modelling of massive stars and supernovae (SNe) plays a crucial role in understanding galaxies. From this modelling we can derive fundamental constraints on stellar evolution, mass-loss processes, mixing, and the products of nucleosynthesis. Proper account must be taken of all important processes that populate and depopulate the levels (collisional excitation, de-excitation, ionization, recombination, photoionization, bound-bound processes). For the analysis of Type Ia SNe and core collapse SNe (Types Ib, Ic and II) Fe group elements are particularly important. Unfortunately little data is currently available and most noticeably absent are the photoionization cross-sections for the Fe-peaks which have high abundances in SNe. Important interactions for both photoionization and electron-impact excitation are calculated using the relativistic Dirac Atomic $R$-matrix Codes ({\sc darc}) for low ionization stages of cobalt. All results are calculated up to photon energies of 45 eV and electron energies up to 20 eV. The wavefunction representation of Co \protect{\sc iii} has been generated using {\sc grasp0} by including the dominant 3d$^7$, 3d$^6$[4s, 4p], 3p$^4$3d$^9$ and 3p$^6$3d$^9$ configurations, resulting in 292 fine structure levels. Electron-impact collision strengths and Maxwellian averaged effective collision strengths across a wide range of astrophysically relevant temperatures are computed for Co \protect{\sc iii}. In addition, statistically weighted level-resolved ground and metastable photoionization cross-sections are presented for Co \protect{\sc ii} and compared directly with existing work.
\end{abstract}




\section{Introduction}\label{sec:introduction}
Lowly ionised species of Cobalt are often observed in astrophysical objects such as supernovae (SNe), cool stars \citep{2010MNRAS.401.1334B}, early type stars \citep{1993A&A...274..335S} and the solar spectrum \citep{1998ApJS..117..261P}. These applications necessitate the need for high quality atomic data which accurately describe the processes of excitation and photoionization. This is further evidenced in SNe by following the nucleosynthesis decay path of $^{56}$Ni$\rightarrow ^{56}$Co$\rightarrow ^{56}$Fe, which occurs post explosion. Our principal aim is to facilitate modelling within the astrophysics community with accurate and up-to-date atomic transitions necessary for synthetic spectral analysis, allowing detailed comparisons to be carried out with observation. Stand alone reports stress both the importance and absence of photon/electron interaction with systems of Iron, Cobalt and Nickel \citep{1995ASPC...78..291R, 2011Ap&SS.336...87H, 2014MNRAS.441.3249D}. 

The Opacity Project has been an invaluable source for such data, but is often limited when considering Fe-peak species. These important Fe-peaks are difficult to investigate due to their open d-shell structure which gives rise to many hundreds of target states for each electronic configuration and typically thousands of closely coupled channels. Hence the target states require substantial configuration interaction expansions for their accurate representation. Fe {\sc ii} is one such challenging case where over the last decade calculations for this ion have grown in size, complexity and sophistication. Significant differences, however, are still observed in the resulting atomic data as can be seen by the latest two major evaluations for the electron-impact excitation of Fe {\sc ii} \citep{2007A&A...475..765R, 2015ApJ...808..174B}. Factors of 2-3 disparity being the norm at the temperature of maxiumum abundance 10$^4$K for many of the low-lying forbidden lines.

There have been a number of studies focused on essential atomic data between species of Co {\sc i}-{\sc iii} concerning bound transitions. These include oscillator strengths for neutral cobalt between 2276 - 9357${\rm \AA}$ \citep{1982ApJ...260..395C}, transition probabilities through a multi configuration approach for comparison with observed infrared spectra \citep{1988A&A...200L..25N}, and also a relativistic Hartree-Fock approach between the lowest 47 levels of Co {\sc ii} \citep{1998A&AS..129..147Q}. More recently, collision strengths and other radiative data have been calculated for Co {\sc ii} \citep{2015arXiv150903164S} and Co {\sc iv} \citep{2015arXiv150907648A}. During the preparation of this work, it has come to our attention a detailed study of electron-impact excitation cross-sections for Co {\sc iii} conducted by \citet{2016arXiv160200712S}.

The early ion stages of, and even neutral Cobalt are clearly important as detailed in the literature. Early observations has shown strong Co {\sc ii} lines in $\eta$ Carinae \citep{1976MNRAS.174P..59T}, confirmed more recently by \citet{2001AJ....122..322Z} to be unusually strong, and in the UV regime in $\zeta$ Oph \citep{1979ApJ...234..506S}, confirmed by \citet{1993ApJ...413L..51F}. These lines are apparent in the binary star HR5049 \citep{1980A&A....85..138D, 1982Obs...102..138D} - which previously have been unidentified due to the lack of laboratory data. In this same study, the amount of Cobalt is estimated to be around 3.0dex overabundant relative to the Sun.

Due to the decay path of $^{56}$Ni, Cobalt is often observed in various SNe at both early and late epochs. The Type {\sc ii} SNe 1987A, exploded in the large magellanic cloud, providing a study of the expanding ejecta as the $^{56}$Co decays. A large proportion of Fe {\sc ii} and Co {\sc ii} lines are blended due to their similar ionization energies, but the strong 1.547$\mu$m line occurs from the transition a$^5$F$_5\rightarrow$ b$^3$F$_4$ \citep{1989MNRAS.238..193M, 1993ApJ...419..824L} in Co {\sc ii}. The Co {\sc iii}  a$^4$F$_{9/2}\rightarrow$ a$^2$G$_{9/2}$ 0.589$\mu$m line in another Type {\sc ii} SNe, 1991bg, is used as a diagnostic to infer the mass of synthesized $^{56}$Ni. It is also possible to deduce important properties such as the mass of the exploding star \citep{1997MNRAS.284..151M}.

The Co {\sc ii} and Co {\sc iii} ions under discussion in this publication have also received much interest over the last decade. The mid infrared spectrum of SNe 2003hv and 2005df show strong Co {\sc iii} line emissions and even emission from Co {\sc iv} \citep{2007ApJ...661..995G}. However, the collisional processes included in the model have been approximated using statistically weighted collision strengths. A study of the near infrared spectra from SNe 2005df yields strong Co {\sc iii} emission lines initially but by day 200 the majority of Cobalt has expectedly decayed down to Fe \citep{2015ApJ...806..107D}. A number of Co {\sc iii} lines are still visible at late times for Type Ia SNe as documented in \citep{1995ASPC...78..291R}. These lines would be extremely beneficial in particular diagnostic work, but as outlined above, little collisional data exists. In addition the photoionization cross sections employed in the models are obtained from a central potential approximation \citep{1979ApJS...40..815R} for the ground state only. Co {\sc iii} is also present in SNe 2014J and the 11.888$\mu$m line is useful for monitoring the time evolution of the photosphere, and again, the mass of synthesized Ni \citep{2015ApJ...798...93T}. It is evident from these works and the associated applications the importance of conducting sophisticated and complete calculations for the lowly ionised Fe-peak species of Fe, Ni and Co.

In Section 2 we discuss the development of an accurate structure model for Co {\sc iii} to include in the $R$-matrix collisional calculations for electron-impact excitation and photoionization. The accuracy of this model will be tested by reassesing energy levels of the target states and the conformity of transition probabilities with previous assessments. In Section 3 we present level-resolved ground and excited state photoionization cross-sections for Co {\sc ii} and a selection of collision strengths and effective collision strengths for the electron impact excitation of Co {\sc iii}. Comparisons will be made where possible with existing data but these are limited. Finally we summarise our findings and conclusions in Section 4.
\section{Important transitions for synthetic spectral modelling}\label{sec:structure}
This paper focuses initially on transitions that occur between discrete states of our atomic system, Co {\sc iii}. In Figure 1 we graphically present some of the most important lines in the infrared and visible energy bands of the spectrum. Transitions among the ground state term $^4$F and levels of the parent ion Co {\sc iii} with configuration 3d$^7$ are shown on the right hand side. The neighbouring system Co {\sc ii} is also shown on the left complete with its fine-structure split $J$ levels to indicate the photoionization process under investigation. 

\subsection{Target description and bound state transitions}
We first detail results obtained from transitions within the bound system of Co {\sc iii} and our {\it ab initio} energy eigenvalues. The Dirac-Coulomb Hamiltonian can be written as,
\begin{equation}\label{eq:dirac}
H_{D}= \sum_{i=1}^N -ic\bm{\alpha} \nabla_i + (\bm{\beta} - 1)c^2-\frac{Z}{r_i} + \sum_{i<j}^N\frac{1}{|\mathbf{r}_j-\mathbf{r}_i|}
\end{equation}
which is then incorporated into the {\sc grasp0} computer package. $\bm{\alpha}$ and $\bm{\beta}$ relates the Pauli spin matrices, $c$ is the speed of light, $Z=27$ is the atomic number and $N=25$ is the number of electrons for the system.

%
\begin{figure}
\includegraphics[scale=0.36, angle=-90]{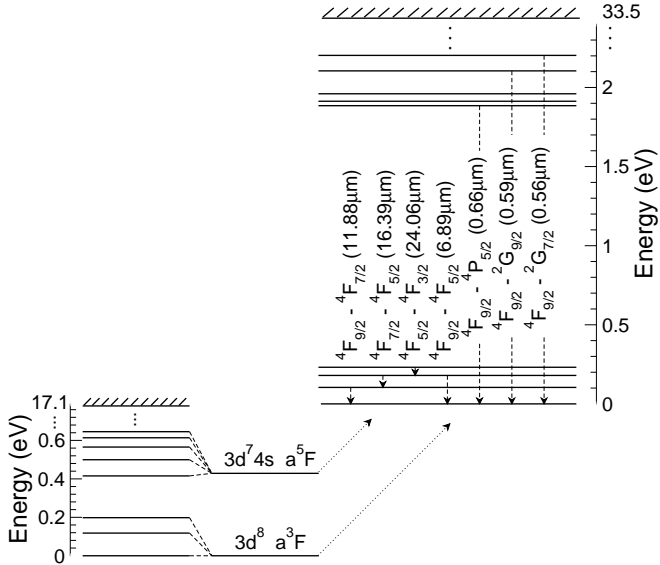}
\caption{Important lines in the infrared and visible energy bands between levels of Co {\sc iii} amongst the 3d$^7$ configuration and involving the ground term $^4$F. The neighbouring system of Co {\sc ii} is to the left with its split $J$ levels to indicate the photoionization process. \label{fig:trans}}
\end{figure}
%

%
\begin{table*}
\begin{center}
\begin{tabular}{@{} l *8c @{}}
            \toprule
            
\multicolumn{1}{c}{Index} & Level & \textit{S\&C} & \textit{model 1}  & \textit{model 2} & $\%$ & \textit{model 3} & \textit{Storey} \\

            \midrule
            
\multicolumn{1}{c}{  1} & 3d$^7$ a$^4$F$_{ 9/2}$ &  0.00000  & 0.00000  & 0.00000 & 0.0 & 0.00000  & 0.00000\\
\multicolumn{1}{c}{  2} & 3d$^7$ a$^4$F$_{ 7/2}$ &  0.10430  & 0.10355  & 0.09939 & 4.7 & 0.09980    & 0.10216\\
\multicolumn{1}{c}{  3} & 3d$^7$ a$^4$F$_{ 5/2}$ &  0.17994  & 0.18001  & 0.17255 &  4.1 & 0.17321   & 0.17705\\
\multicolumn{1}{c}{  4} & 3d$^7$ a$^4$F$_{ 3/2}$ &  0.23145  & 0.23263  & 0.22280 &  3.7 & 0.22362  & 0.22838 \\
\multicolumn{1}{c}{  5} & 3d$^7$ a$^4$P$_{ 5/2}$ &  1.88480  &  2.39405  & 2.20436 &  16.9 & 2.19761   &  2.29369\\
\multicolumn{1}{c}{  6} & 3d$^7$ a$^4$P$_{ 3/2}$ &  1.91285  & 2.42637  & 2.23271 &   16.7 & 2.22608   & 2.32904\\
\multicolumn{1}{c}{  7} & 3d$^7$ a$^4$P$_{ 1/2}$ &  1.96036  & 2.47283 & 2.27875 &   16.2 & 2.27244   &   2.37033\\
\multicolumn{1}{c}{  8} & 3d$^7$ a$^2$G$_{ 9/2}$ &  2.10496  & 2.39059  & 2.38992 &  13.5 & 2.38842   & 2.42773\\
\multicolumn{1}{c}{  9} & 3d$^7$ a$^2$G$_{ 7/2}$ &  2.20273  & 2.48967  & 2.48250 &   12.7 & 2.48138 & 2.52395 \\
\multicolumn{1}{c}{ 10} & 3d$^7$ a$^2$P$_{ 3/2}$ &  2.50385   & 3.16229 &     2.85243 &   13.9 & 2.84334   & 3.17809\\
\multicolumn{1}{c}{ 11} & 3d$^7$ a$^2$P$_{ 1/2}$ &  2.59356  & 3.27065 &    2.96428 & 14.3 & 2.95673    & 3.28236 \\
\multicolumn{1}{c}{ 12} & 3d$^7$ a$^2$H$_{11/2}$ &  2.81696  & 3.18356 &    3.35270 & 19.0 & 3.35260    & 3.18478\\
\multicolumn{1}{c}{ 13} & 3d$^7$ a$^2$H$_{ 9/2}$ &  2.90548  &  3.26836 &     3.43116 &  18.1 & 3.43143  & 3.27058\\
\multicolumn{1}{c}{ 14} & 3d$^7$ a$^2_2$D$_{5/2}$ &  2.85893  & 3.44672 &    3.06277 &  7.1 & 3.04740   & 3.44726\\
\multicolumn{1}{c}{ 15} & 3d$^7$ a$^2_2$D$_{ 3/2}$ &  3.00498  & 3.60358 &    3.23140 & 7.5 & 3.21873   &  3.59963 \\
\multicolumn{1}{c}{ 16} & 3d$^7$ a$^2$F$_{ 5/2}$ &  4.59002  &  5.51392 &    5.36609 &  16.9 & 5.35716   & - \\
\multicolumn{1}{c}{ 17} & 3d$^7$ a$^2$F$_{ 7/2}$ &  4.62666  &  5.55837 &    5.40863 &   16.9 & 5.39998  & - \\
\multicolumn{1}{c}{ 18} & 3d$^6$4s a$^6$D$_{ 9/2}$ &  5.75762  & 6.70846 &     6.06849 & 5.4 & 6.20178   & - \\
\multicolumn{1}{c}{ 19} & 3d$^6$4s a$^6$D$_{ 7/2}$ &  5.82764  &  6.78942 &     6.14650 &  5.5 & 6.27925 & -  \\
\multicolumn{1}{c}{ 20} & 3d$^6$4s a$^6$D$_{ 5/2}$ &  5.87876  &  6.84869 &    6.20378 & 5.5 & 6.33614   & - \\
\multicolumn{1}{c}{ 21} & 3d$^6$4s a$^6$D$_{ 3/2}$ &  5.91387  & 6.88947 &    6.24324 &  5.5 & 6.37536 & - \\ 
\multicolumn{1}{c}{ 22} & 3d$^6$4s a$^6$D$_{ 1/2}$ &  5.93448  & 6.91340 &    6.26643 &  5.6 & 6.39840 & - \\
\multicolumn{1}{c}{ 23} & 3d$^6$4s a$^4$D$_{ 7/2}$ &  6.90954  & 8.42015 &    7.72255 &  11.8 & 8.04998   & - \\
\multicolumn{1}{c}{ 24} & 3d$^6$4s a$^4$D$_{ 5/2}$ &  6.98946  &  8.51285 &    7.81196 &   11.8 & 8.13868& - \\
\multicolumn{1}{c}{ 25} & 3d$^6$4s a$^4$D$_{ 3/2}$ &  7.04166  &  8.57368 &    7.87084 &  11.8 & 8.19722 & - \\
\multicolumn{1}{c}{ 26} & 3d$^6$4s a$^4$D$_{ 1/2}$ &  7.07167  &  8.60877 &    7.90480 &  11.8 & 8.23095 & - \\
\multicolumn{1}{c}{ 27} & 3d$^7$ a$^2_1$D$_{ 3/2}$ &  -  &  8.67161 & 7.93064 &   - & 7.96545 & - \\
\multicolumn{1}{c}{ 28} & 3d$^7$ a$^2_1$D$_{ 5/2}$ &  -  & 8.61018 &  8.00119 & - & 7.89375 & - \\
\multicolumn{1}{c}{ 29} & 3d$^6$4s b$^4$P$_{ 5/2}$ &  8.79471  &  10.25065 &   9.63834 &  9.6 & 9.80974  &   - \\
\multicolumn{1}{c}{ 30} & 3d$^6$4s a$^4$H$_{13/2}$ &  8.88014  &  10.00120 &    9.38243 &  5.6 & 9.55349  & - \\
\multicolumn{1}{c}{ 31} & 3d$^6$4s a$^4$H$_{11/2}$ &  8.91121  & 10.03198  &    9.41243 &  5.6 & 9.58337  & - \\
\multicolumn{1}{c}{ 32} & 3d$^6$4s a$^4$H$_{ 9/2}$ &  8.93719  &  10.05805 &    9.43786 &  5.6 & 9.60883   & - \\
\multicolumn{1}{c}{ 33} & 3d$^6$4s a$^4$H$_{ 7/2}$ &  8.96040  &  10.08057 &    9.45963 &  5.6 & 9.63075   & - \\
\multicolumn{1}{c}{ 34} & 3d$^6$4s b$^4$P$_{ 3/2}$ &  8.96926  &  10.46563 &   9.84580 &  9.8 & 10.01710   & - \\
\multicolumn{1}{c}{ 35} & 3d$^6$4s b$^4$P$_{ 1/2}$ &  9.07744  &  10.59439 &    9.97061 &  9.8 & 10.14009  & - \\
\multicolumn{1}{c}{ 36} & 3d$^6$4s b$^4$F$_{ 9/2}$ &  9.08631  &  10.42302 &    9.81043 &  8.0 & 9.98184  & - \\
\multicolumn{1}{c}{ 37} & 3d$^6$4s b$^4$F$_{ 7/2}$ &  9.11782  &  10.46016 &    9.84605 &  8.0 & 10.01692  & - \\
\multicolumn{1}{c}{ 38} & 3d$^6$4s b$^4$F$_{ 5/2}$ &  9.14093  &  10.49101 &    9.87566 &  8.0 & 10.04624  & - \\
\multicolumn{1}{c}{ 39} & 3d$^6$4s b$^4$F$_{ 3/2}$ &  9.15770  &  10.51540 &    9.89893 &  8.1 & 10.06929  & - \\
\multicolumn{1}{c}{ 40} & 3d$^6$4s a$^4$G$_{11/2}$ &  9.48714  & 10.77237 &    10.15994 &   7.1 & 10.33533 & - \\
\multicolumn{1}{c}{ 41} & 3d$^6$4s b$^2$P$_{ 3/2}$ &  9.52088  & 11.32810 &   10.67964 &   12.2 & 10.96474 & - \\
\multicolumn{1}{c}{ 42} & 3d$^6$4s a$^4$G$_{ 9/2}$ &  9.56180  & 10.85246 &    10.23781 & 7.1 & 10.40898   & - \\
\multicolumn{1}{c}{ 43} & 3d$^6$4s a$^4$G$_{ 7/2}$ &  9.59428  & 10.88642 &    10.27110 &  7.0 & 10.44158  & - \\
\multicolumn{1}{c}{ 44} & 3d$^6$4s b$^2$H$_{11/2}$ &  9.59782  & 11.04902 &    10.39701 &  8.3 & 10.67836  & - \\
\multicolumn{1}{c}{ 45} & 3d$^6$4s a$^4$G$_{ 5/2}$ &  9.60534  & 10.89526 &    10.28019 &  7.0 & 10.45125  & - \\
\multicolumn{1}{c}{ 46} & 3d$^6$4s b$^2$P$_{ 1/2}$ &  9.72461  & 11.57042 &    10.91582 &  12.2 & 11.19712   & - \\
\multicolumn{1}{c}{ 47} & 3d$^6$4s b$^2$H$_{ 9/2}$ &  9.62402  & 11.08283 &   10.42907 &  8.4 & 10.71376  & - \\
\multicolumn{1}{c}{ 48} & 3d$^6$4s b$^2$F$_{ 7/2}$ &  9.78580  & 11.44778 &    10.80218 &  10.4 & 11.08548   & - \\
\multicolumn{1}{c}{ 49} & 3d$^6$4s b$^2$F$_{ 5/2}$ &  9.84749  & 11.53372 &   10.88286 &   10.5 & 11.16801 & - \\
\multicolumn{1}{c}{ 50} & 3d$^6$4s b$^2$G$_{ 9/2}$ & 10.21175 & 11.83144 &   11.18470 &  9.6 & 11.47076 & - \\

            \bottomrule
            
 \end{tabular}
 \caption{Energies for the lowest 50 levels of Co {\sc iii} are presented in eV relative to the ground state 3d$^7$ a$^4$F$_{9/2}$. \textit{S\&C} is from the work of \citet{1985aeli.book.....S}. \textit{model 1}, \textit{model 2}, \textit{model 3} are the current results from {\sc grasp}0 and the last column are the lowest 15 levels from \citet{2016arXiv160200712S}. We also present the $\%$ difference between our current \textit{model 2} and \textit{S\&C} in the 6th column. \label{tab:elevels}}
 \end{center}
\end{table*}
%

As stated in the introduction, partially filled d-shell systems are difficult due to the hundreds of levels associated with a single configuration. Initially, we include three configurations, 3d$^7$ and 3d$^6$[4s, 4p] during the optimization process, denoted as \textit{model 1}, which results in a total of 262 fine structure levels to describe the Co {\sc iii} ion. The ground state of which is 3d$^7$a$^4$F$_{9/2}$. Next we optimize all orbitals up to 3d on the configurations from the double electron promotions, 3s$^2$, 3p$^2$ $\rightarrow$ 3d$^2$ and include in the total calculation all configurations from \textit{model 1} plus 3p$^6$3d$^9$ (double promotion from 3s to 3d) and 3s$^2$3p$^4$3d$^9$ (double promotion from 3p to 3d). This technique can be useful as it alleviates the necessity to include numerous pseudo states into the calculation. We denote this \textit{model 2} which constitutes a total of 292 levels. Finally, by including 3d$^5$[4s$^2$, 4p$^2$] and 3d$^5$4s4p, the number of levels drastically increases to 1,259 levels, and we label this as \textit{model 3}. 

We present in Table \ref{tab:elevels} our {\it ab initio} energy levels obtained from {\sc grasp}0 in eV for the lowest lying 50 levels alongside those observed by \citet{1985aeli.book.....S}. We also present the $\%$ difference between \citet{1985aeli.book.....S} and our \textit{model 2}, and also provide the lowest 15 levels of \citet{2016arXiv160200712S}. Good agreement is found ($<$10\% for the majority of levels) between the present \textit{model 2} and \textit{model 3} energies and those of \citet{2016arXiv160200712S}. As with other Fe-peak ions, the energy levels of the lowest-lying 3d$^{7}$ fine-structure states are notoriously difficult to determine. The highest disparities are found for these levels when compared with \citet{1985aeli.book.....S}, the largest being for the 3d$^{7}$ $^{4}$F$_{9/2}$ $\rightarrow$ 3d$^{7}$ $^{2}$H$_{11/2}$ (1-12) transition. For levels indexed above 17 the differences are at most 10-11$\%$ and for many levels by considerable less. The main problem is due to the fact that a single 3d orbital is used to describe the configuration state functions for multiple configurations of type 3d$^{7}$ and 3d$^{6}$4s. Similar differences were reported by \citet{2009ADNDT..95..910R} for the low-lying 3d$^{7}$ fine-structure levels of Fe {\sc ii}. 
Despite the high $\%$ differences found between these lowest levels in Table \ref{tab:elevels}, overall the average $\%$ change across the 171 \citet{1985aeli.book.....S} J$\pi$ symmetries is a more acceptable 6.2$\%$. The differences between \textit{model 2} and \textit{model 3} are not significant enough to justify the much larger calculation, and we can also benefit by employing all 292 levels into the close-coupling wavefunction expansion of the Co {\sc iii}. We therefore adopt our \textit{model 2} as the final model for the scattering calculation. 

\begin{figure}
\includegraphics[scale=0.31, angle=-90]{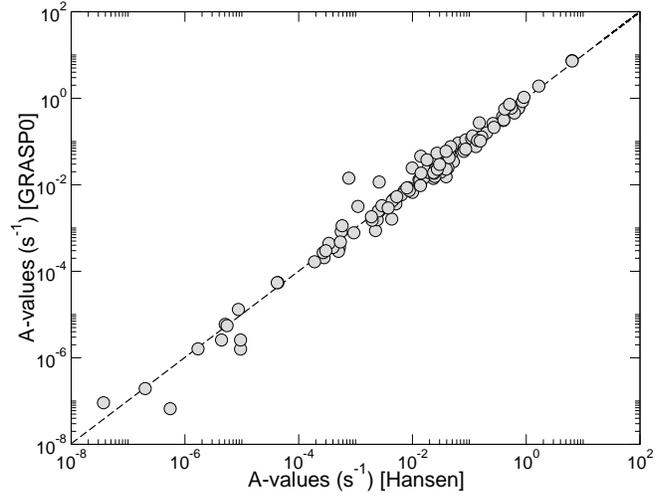}
\caption{Present theoretical $A$-values after the scaling factors in equation (\ref{eq:deltaE}) have been applied, plotted against the results of \citet{1984ApJ...277..435H} in s$^{-1}$ on a log/log scale. \label{fig:avalues}}
\end{figure}

The $A$-value, or transition probability, can be calculated at this stage of the calculation. It is a direct measure of the line strength between two states of our system, and we therefore require accurate wavefunctions and energies. We can account for the discrepancy between {\sc grasp}0 and \citet{1985aeli.book.....S} energy states by considering the following multiplicative scaling factor,
\begin{equation}\label{eq:deltaE}
\Bigg(\frac{\Delta E^{{\rm expt}}_{ji} }{\Delta E^{{\rm theo}}_{ji}}\Bigg)^{\eta},
\end{equation}
Here we set $\eta=3$ or $\eta=5$ for electric and magnetic dipole or quadrupole transitions between two states $j$ and $i$. It is found that these are shifted by a fraction of the recorded values.
A procedure through a least squares fitting process \citep{1981tass.book.....C} is the most commonly implemented source of $A$-values for this ion stage of Cobalt \citep{1984ApJ...277..435H} to date. A total of 130 forbidden transitions within this 3d$^7$ complex are reported, and are provided for comparison in Figure \ref{fig:avalues} against our results after the scaling in equation (\ref{eq:deltaE}) has been performed. The $A$-values are presented on a logarithmic vs. logarithmic scale to incorporate the various magnitudes of results. To investigate this comparison more closely, we present in Table \ref{tab:avalues} a selection of transitions among the lowest 7 levels of Co {\sc iii}. Comparisons are made between the values of \citep{1984ApJ...277..435H}, the present \textit{model 2} with scaled transition energy levels and the recent results from \citet{2016arXiv160200712S}. Excellent agreement is evident for the bulk of the transitions considered with the greatest difference of $12.7\%$ occurring for the 3d$^7$a$^4$F$_{3/2}$ - 3d$^7$a$^4$P$_{5/2}$ (4-5) transition.

%
%

%
\begin{table}
\begin{center}
\begin{tabular}{@{} l *5c @{}}
      \toprule
\multicolumn{1}{c}{ $i - j$}  & \textit{Fivet}  &  \textit{Hansen}   &  \textit{model 2}    &  \textit{Storey}  \\   

      \midrule

\multicolumn{1}{c}{  1 -- 2}  & 2.00$\times 10^{-2}$    & 2.00$\times 10^{-2}$    &   2.00$\times 10^{-2}$  &      2.00$\times 10^{-2}$ \\
\multicolumn{1}{c}{  1 -- 3}  & - & 1.80$\times 10^{-9}$ &  1.75$\times 10^{-10}$   & -  \\
\multicolumn{1}{c}{  1 -- 5}  & 6.65$\times 10^{-2}$    & 4.80$\times 10^{-2}$  &  4.53$\times 10^{-2}$   &  5.55$\times 10^{-2}$  \\
\multicolumn{1}{c}{  2 -- 3}  &  1.31$\times 10^{-2}$    &1.30$\times 10^{-2}$ &  1.31$\times 10^{-2}$   & 1.31$\times 10^{-2}$ \\
\multicolumn{1}{c}{  2 -- 4}  &   -   &5.90$\times 10^{-10}$  &  5.70$\times 10^{-10}$ &  -   \\
\multicolumn{1}{c}{  2 --  5}  &1.78$\times 10^{-2}$    & 1.35$\times 10^{-2}$    &  1.26$\times 10^{-2}$  &     1.51$\times 10^{-2}$ \\
\multicolumn{1}{c}{  2  -- 6}  & 3.73$\times 10^{-2}$    & 2.70$\times 10^{-2}$     &  2.58$\times 10^{-2}$  &   3.14$\times 10^{-2}$  \\
\multicolumn{1}{c}{  3  -- 4}  &4.63$\times 10^{-2}$    &  4.70$\times 10^{-3}$   &  4.63$\times 10^{-3}$  &   4.63$\times 10^{-3}$  \\
\multicolumn{1}{c}{  3  -- 5} &   -     &  2.60$\times 10^{-3}$   &   2.45$\times 10^{-3}$  &    3.14$\times 10^{-3}$  \\
\multicolumn{1}{c}{  3  -- 6}  &2.21$\times 10^{-2}$    &  1.63$\times 10^{-2}$  &    1.50$\times 10^{-2}$  &     1.85$\times 10^{-2}$   \\
\multicolumn{1}{c}{  3 --  7}  &2.73$\times 10^{-2}$    &  2.00$\times 10^{-2}$    &   1.89$\times 10^{-2}$  &     2.30$\times 10^{-2}$ \\
\multicolumn{1}{c}{  4 --  5}  & -    &  4.00$\times 10^{-4}$  &  3.55$\times 10^{-4}$   &   - \\
\multicolumn{1}{c}{  4  -- 6}  & -    &  4.40$\times 10^{-3}$     &  4.22$\times 10^{-1}$  &  5.14$\times 10^{-3}$ \\
\multicolumn{1}{c}{  4 --  7}  & 3.60$\times 10^{-2}$    &  2.60$\times 10^{-2}$   &  2.47$\times 10^{-2}$  &    3.02$\times 10^{-2}$  \\
\multicolumn{1}{c}{  5 --  6}  & -    &  2.70$\times 10^{-4}$  &  2.69$\times 10^{-4}$  &  -   \\
\multicolumn{1}{c}{  5 --  7}  & -   &  5.50$\times 10^{-9}$    &  5.20$\times 10^{-9}$  &   -\\
\multicolumn{1}{c}{  6 -- 7} & -     &  2.50$\times 10^{-3}$    &  2.47$\times 10^{-3}$  &   2.45$\times 10^{-3}$ \\

      \bottomrule
 \end{tabular}
 \caption{$A$-values from \citet{2016A&A...585A.121F}, \citet{1984ApJ...277..435H}, our current \textit{model 2} after the multiplicative scaling factors in equation (\ref{eq:deltaE}) have been applied and results from Storey for transitions amongst the lowest 7 levels.  \label{tab:avalues}}
 \end{center}
 \end{table}
%

A much larger calculation for doubly ionized Fe-peak species has been performed recently by \citet{2016A&A...585A.121F} using the same suite of codes as \citet{1984ApJ...277..435H}, and also by considering the computer package {\sc autostructure}, \citep{1974CoPhC...8..270E, 1986JPhB...19.3827B}, where the optimization process is carried out with a Thomas-Fermi-Dirac potential using lambda scaling parameters. Numerous doubly ionized ions of Fe, Ni and Co were considered in this publication. Single and double electron promotions out of the 3d, and also single electron promotions out of the 3s to the 5s orbital were included and comprised the basis expansion for both calculations. This is the most sophisticated and complete report for radiative rates in Co {\sc iii} to date. Within the 3d$^7$ complex, we vary approximately $27\%$ on average compared with both methods of \citet{2016A&A...585A.121F}.

\subsection{The $R$-matrix method}
To extend this problem to include interactions with photons and electrons, we consider the $R$-matrix method. A general overview of the theory can be found in \citet{2011rmta.book.....B}. The theory was developed to perform electron scattering by \citet{1971JPhB....4..153B} and extended for photoionization by \citet{1975JPhB....8.2620B}.

These calculations can be performed using a non orthogonal B-spline basis set approach \citep{2000JPhB...33..313Z, 2004JPhB...37.4693Z}, intermediate-coupling frame transformation methods \citep{1998JPhB...31.3713G} and one-body perturbation corrections to the non-relativistic Hamiltonian via the Breit-Pauli operators \citep{1980JPhB...13.4299S} to name a few. In this paper, we consider the Dirac-Atomic $R$-matrix codes ({\sc darc}) to accurately include important relativistic effects through the Dirac Hamiltonian in equation (\ref{eq:dirac}) for this mid to heavy species ion. The wavefunctions obtained from {\sc grasp0} are formatted suitable for inclusion into {\sc darc}. Validity of these methods between the outlined theoretical approaches are always important, and enhance the robustness of the techniques carried out as seen in \citet{2016MNRAS.456..366T} for Ar$^+$ and \citet{2015MNRAS.450.4174F} for Al$^{9+}$. Therefore, it is possible to benchmark sophisticated experimental techniques against theory \citep{2009JPhB...42w5602M, 2011JPhB...44q5208G}.
	
\subsection{Photoionization}
The photoionization process is described by,
\begin{equation}\label{eq:photo}
h\nu + {\rm Co}^{+} ~ \{{\rm 3d}^8 {\rm a}^3{\rm F}^{{\rm e}}_J, {\rm 3d}^7 {\rm 4s~a}^5{\rm F}^{{\rm e}}_J\} \rightarrow {\rm Co}^{2+} + e^{{\rm -}}
\end{equation}
where a photon leads to the ionization of an electon directly, or via a Rydberg resonance. The process in equation (\ref{eq:photo}) can be formally calculated as,
\[
\sigma^{{\rm p}}_{i \rightarrow j} = \frac{4\pi a_0^2 \alpha \omega}{3g_i}\sum(\Psi_j || \mathbf{D} || \Psi_i)
\]

%
\begin{figure*}
\includegraphics[scale=0.68, angle=-90]{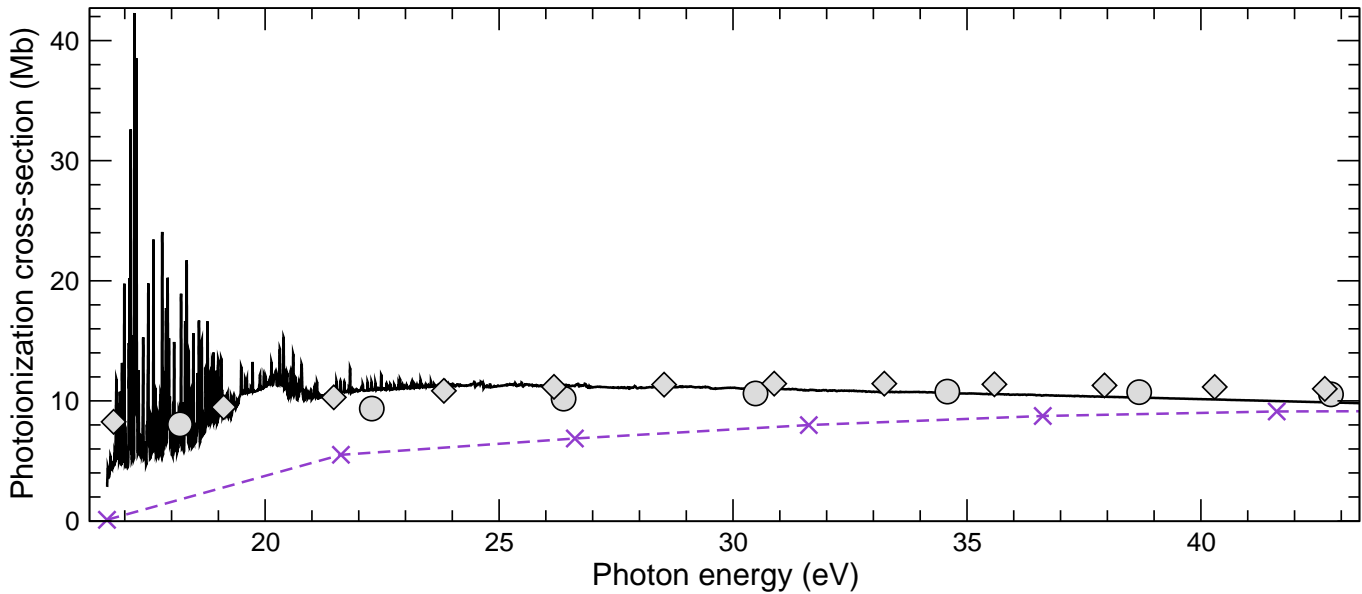}
\caption{Photoionization cross-sections in Mb against the photon energy in eV. The solid black curve is the total initial ground state, statistically weighted 3d$^8$ $^3$F to all allowed dipole final states. The crosses are results from \citet{1979ApJS...40..815R}, the diamonds are from \citet{1993ADNDT..55..233V}, and finally the circles are those from the distorted wave calculation. \citep{2015JPhB...48n4014F} \label{fig:photo1}}
\end{figure*}
%

for initial and final scattering wavefunctions subject to the total dipole contribution. $\omega = h\nu$ is the photon energy or $\omega \rightarrow \omega^{-1}$ when considering the velocity operator, $\mathbf{D}$. $\alpha$ is the fine structure constant, $a_0$ is the bohr radius and $g_i$ is the statistical weight of the initial state wavefunction. The summation is then carried out over all dipole-allowed transitions. The total wavefunctions $\Psi$ are obtained from the momenta couplings with those wavefunctions obtained in Section 2.1. We represent these $\sigma^{{\rm p}}$ in units of Mb ($\equiv 10^{-18}$cm$^2$) throughout this work.

The transitions of interest are calculated within the lowest eight fine-structure levels of Co {\sc ii} pertaining to the two $LS\pi$ states as noted in equation (\ref{eq:photo}). We also maintain a closed 3p$^6$ core and are only concerned with low energy transitions above threshold. Therefore, all even and odd allowed dipole symmetries up to $J = 6$ are calculated subject to the selection rules.

%
\begin{table}
\begin{center}
\begin{tabular}{@{} l *5c @{}}
 \toprule
\multicolumn{1}{c}{Index} & Level & \textit{Pickering} & \textit{Current} & $\Delta$ \\

 \midrule
 
\multicolumn{1}{c}{  1} & 3p$^6$3d$^8$ a$^3$F$_{ 4}$ & -17.0844 & -16.6232 & 0.46 \\
\multicolumn{1}{c}{  2} & 3p$^6$3d$^8$ a$^3$F$_{ 3}$ & -16.9666 & -16.5004 & 0.47\\
\multicolumn{1}{c}{  3} & 3p$^6$3d$^8$ a$^3$F$_{ 2}$ & -16.8864  & -16.4159 & 0.47\\
\multicolumn{1}{c}{  4} & 3p$^6$3d$^7$($^4$F)4s a$^5$F$_{ 5}$ & -16.6690 & -16.3531 & 0.32 \\
\multicolumn{1}{c}{  5} & 3p$^6$3d$^7$($^4$F)4s a$^5$F$_{ 4}$ & -16.5849 & -16.2694 & 0.32\\
\multicolumn{1}{c}{  6} & 3p$^6$3d$^7$($^4$F)4s a$^5$F$_{ 3}$ & -16.5189 & -16.2031 & 0.32\\
\multicolumn{1}{c}{  7} & 3p$^6$3d$^7$($^4$F)4s a$^5$F$_{ 2}$ & -16.4707 & -16.1544 & 0.32\\
\multicolumn{1}{c}{  8} & 3p$^6$3d$^7$($^4$F)4s a$^5$F$_{ 1}$ & -16.4391 & -16.1224 & 0.32\\
 \bottomrule
 
 \end{tabular}
 \caption{Bound state energies of the lowest eight states of Co {\sc ii} relative to the ground state 3d$^7$ a$^3$F$_{9/2}$ of Co {\sc iii} compared with the relative energies of \citet{1998ApJS..117..261P}, labelled as \textit{Pickering}. \label{tab:blevels}}
 \end{center}
\end{table}
%

%
\begin{figure*}
\includegraphics[scale=0.68, angle=-90]{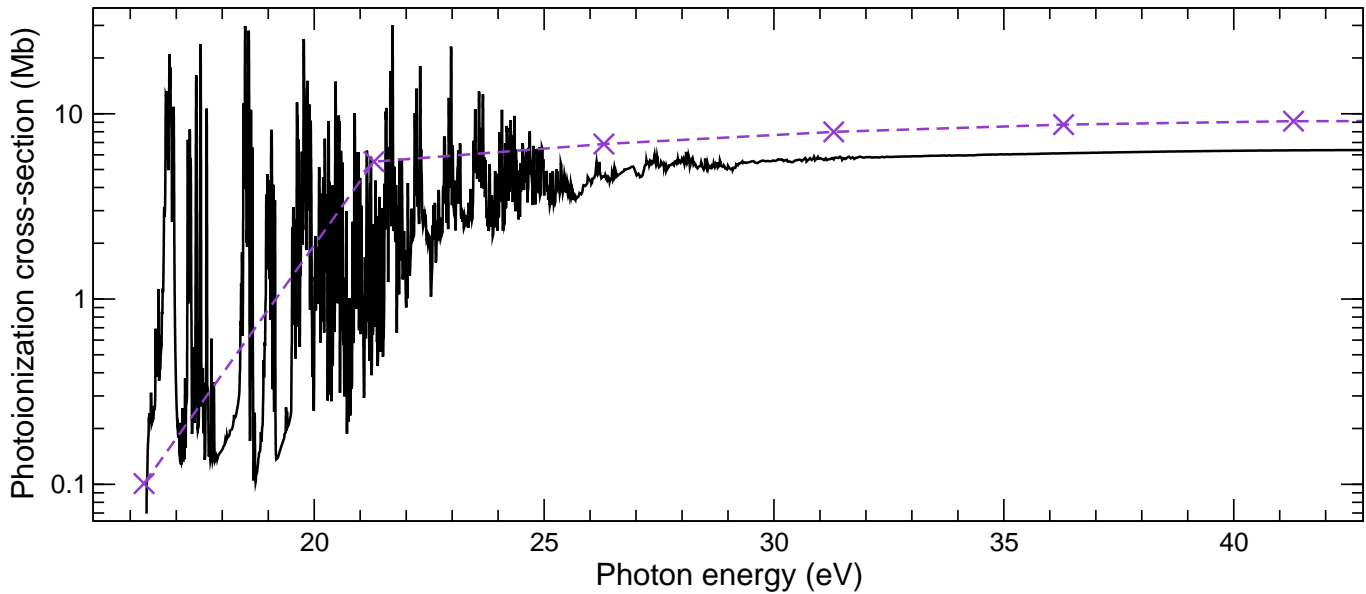}
\caption{Photoionization cross-sections in Mb against the photon energy in eV. The solid black curve is the photoionization, statistically averaged levels of 3d$^7$4s $^5$F to all allowed dipole final states. The crosses are results from \citet{1979ApJS...40..815R}. \label{fig:excited}}
\end{figure*}
%

\subsection{Electron-impact excitation}
The collision strength between an initial state $i$ and final state $j$ can be obtained from the cross section, $\sigma^{e}$, 
\begin{equation}\label{eq:coll}
\Omega_{i\rightarrow j} = \frac{g_ik_i^2}{\pi a_0^2}\sigma^{e}_{j\rightarrow i}.
\end{equation}
These collision strengths represent a detailed spectrum, complete with the already mentioned autoionizing states. To present the results in a more concise format, we assume a Maxwellian distribution of the colliding electron velocities. We define the effective collision strength as
\begin{equation}\label{eq:ups}
\Upsilon_{i \rightarrow j} = \int^{\infty}_{0} \Omega_{i \rightarrow j}e^{-\epsilon_f/kT}d\Big(\frac{\epsilon_f}{kT}\Big),
\end{equation}
where $\epsilon_f$ is the ejected energy of the electron, $3,800 \leq T \leq 40,000$ in K, and $k = 8.617\times 10^{-5}$ eV/K is Boltzmanns constant. Each $\Upsilon_{i \rightarrow j} $ is calculated for 11 electron temperatures.

We calculate the main contribution to the collision strength defined in equation (\ref{eq:coll}) from the partial waves up to $J = 13$ of both even and odd parity. These are obtained by considering appropriate total multiplicity and orbital angular momentum partial waves. However, in order to converge transitions at higher energies, we explicitly calculate partial waves up to $J=38$ and use top-up procedures outlined in \citet{1992A&A...254..436B} to account for further contribution to the total cross-section

\section{Results}
As mentioned previously, the photoionization of Co {\sc ii} and electron-impact excitation of Co {\sc iii} rely on an accurate description of the Co {\sc iii} wavefunctions. We are able to retain consistency throughout the calculation, and the fundamental $R$-matrix conditions apply to both processes. We select a total of 14 continuum basis orbitals per angular momenta to describe the scattered electron. The $R$-matrix boundary is then set at 10.88 a.u. in order to enclose the radial extent of the 4p orbital. To make comparisons between observation, the target thresholds obtained in Table 1 have been shifted to those of \citet{1985aeli.book.....S}. 

\subsection{Photoionization}
In this section, we detail our results from the photoionization process described by equation (\ref{eq:photo}). There is minimal atomic data for this interaction in the literature as only the total ground state transition exists. The first compilation is from \citet{1979ApJS...40..815R}, using Hartree-Slater wavefunctions from a Central field potential. Second, and more recently, \citet{1993ADNDT..55..233V} have calculated cross-sections using the Hartree-Dirac-Slater potential and then applying an analytic fitting procedure. Recently, the Los Alamos suite of codes by \citet{2015JPhB...48n4014F} have been used to obtain results by a distorted wave method. For this we look at the photoionization of a 3d electron into the configuration averaged final states. 
To carefully resolve these spectra, we have employed a total of 200,000 equally spaced energy points over a photoelectron energy range of 20 eV. This ensures the high $nl$ Rydberg resonant states are properly delineated.

The initial bound states that are required, corresponding to the left hand side of equation (\ref{eq:photo}) are calculated first. In Table \ref{tab:blevels} we present the energies relative to the ground state 3d$^7$ a$^3$F$_{9/2}$ of Co {\sc iii} from our current $R$-matrix results and compare with those of \citet{1998ApJS..117..261P}. We deviate $\approx 0.47$ eV for the 3d$^8$ and $\approx 0.32$ eV for the 3d$^7$($^4$F)4s. Despite these discrepancies
good agreement is evident for the splitting between all eight fine-structure levels.

In Figure \ref{fig:photo1} we present the photoionization cross-section representing the statistically weighted initial state 3d$^8$ a$^3$F to all allowed final states. The results in this figure have been convoluted with a 10 meV Gaussian profile at full-width half-maximum to better compare with experiment. We directly compare here with the results of all three previously mentioned theoretical methods \citep{1979ApJS...40..815R, 1993ADNDT..55..233V, 2015JPhB...48n4014F}. These methods do not include autoionization states and therefore only background cross-sections are presented. It is clear however that the results are in excellent agreement, with only \citet{1979ApJS...40..815R} reaching factors of two or more difference in the $< 20$ eV region. As over 200 eigenenergies obtained from {\sc grasp}0 are $<$ 28 eV of photon energy, this above threshold region is dominated by multiple Rydberg resonances series converging onto these states. 

The second statistically weighted bound state a$^5$F of Co {\sc ii} is from the configuration 3d$^7$4s. In Figure \ref{fig:excited} we present the total photoionization cross-section for this metastable level and compare with the earlier work of \citet{1979ApJS...40..815R}. 
This cross-section is weighted as a consequence of its five fine-structure split levels $1 < J < 5$. Again, the photoionization of the 3d electron is favourable, so we expect the majority of the spectrum to be accounted for from the 3d$^6$4s target states, which are accessible at 5.75762 eV above the ionization threshold. Good agreement is found between the two calculations for the background cross section, particularly in the higher photon energy range above 25 eV.

\subsection{Electron-impact excitation}
The procedures outlined in Section 2.4 are now applied to obtain accurate collision strengths and the corresponding effective collision strengths describing the electron-impact excitation of Co {\sc iii}. There has been work carried out recently on both singly ionized Co {\sc ii} \citep{2015arXiv150903164S} and also Co {\sc iii} \citep{2016arXiv160200712S}. The work of \citet{2016arXiv160200712S} is similar to our current evaluation, but differences are evident in the method considered. Their basis set has been optimized using {\sc autostructure} , and also includes a $4\bar{d}$ orbital plus additional configuration interaction. However, only a total of 109 fine structure levels have been included in the close coupling target representation. The semi-relativistic Breit-Pauli $R$-matrix approach was considered during the scattering calculation.

%
\begin{figure}
~\\
~\\
  \subfloat{\includegraphics[scale=0.31, angle=-90]{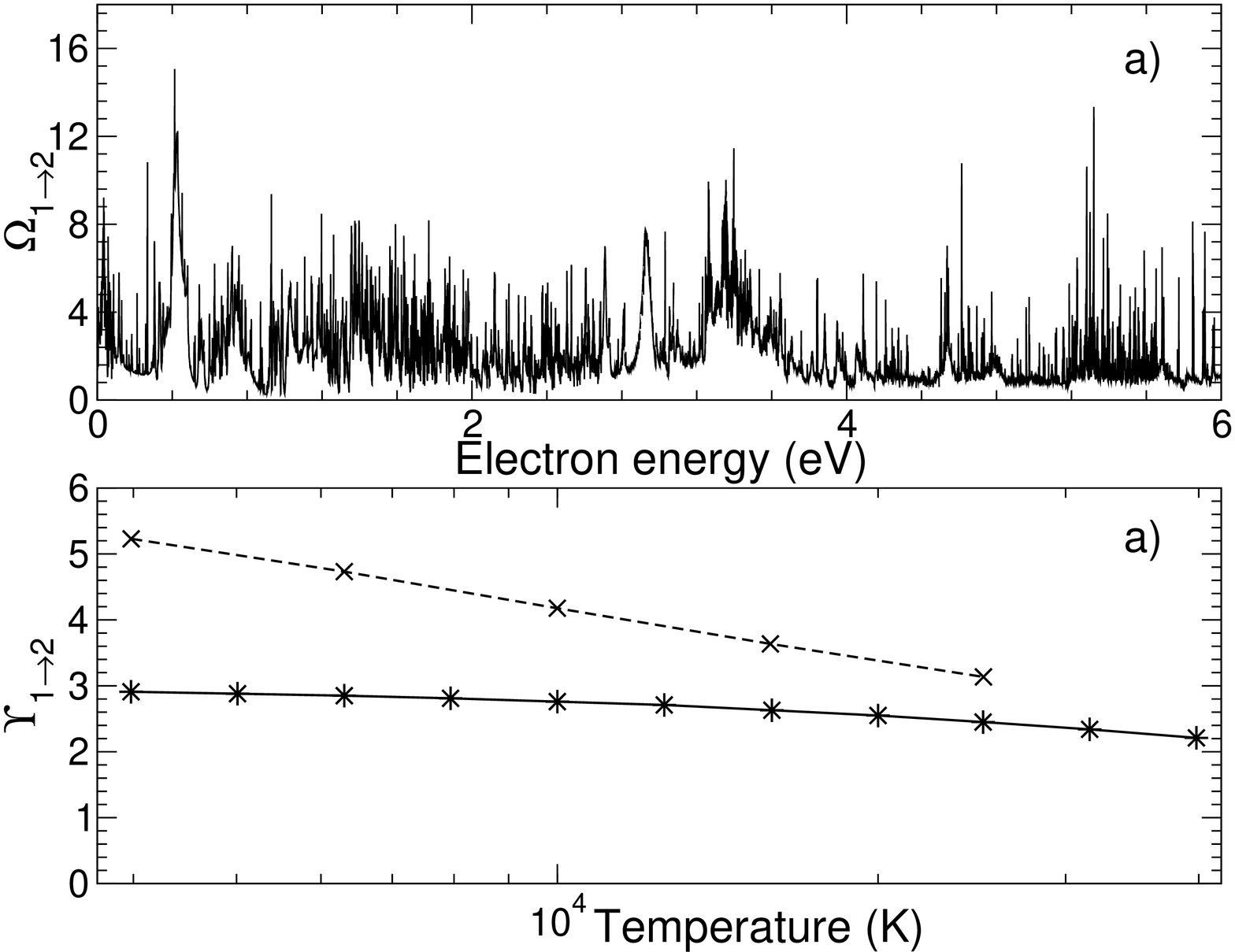}}\\
     \subfloat{\includegraphics[scale=0.31, angle=-90]{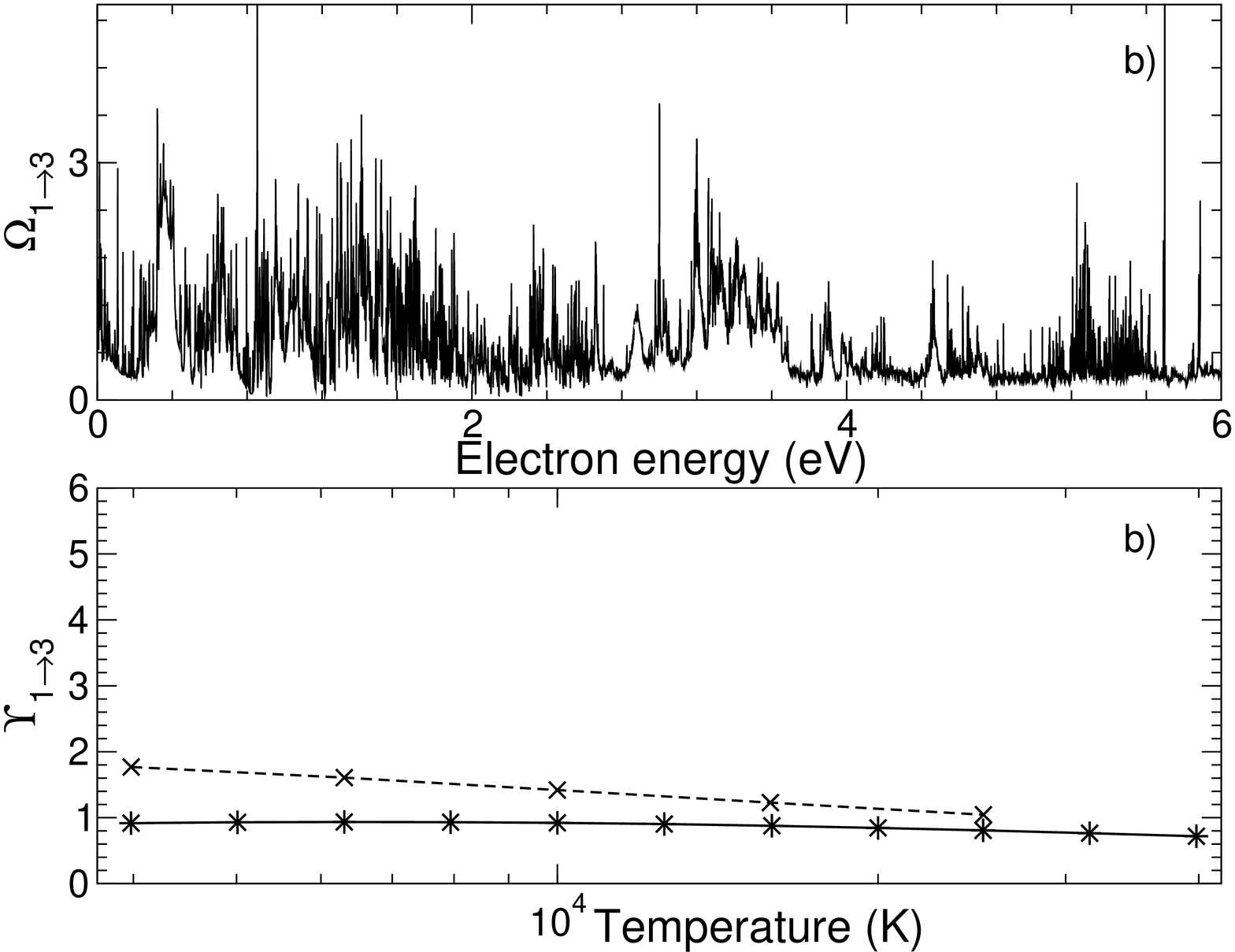}}\\
        \subfloat{\includegraphics[scale=0.31, angle=-90]{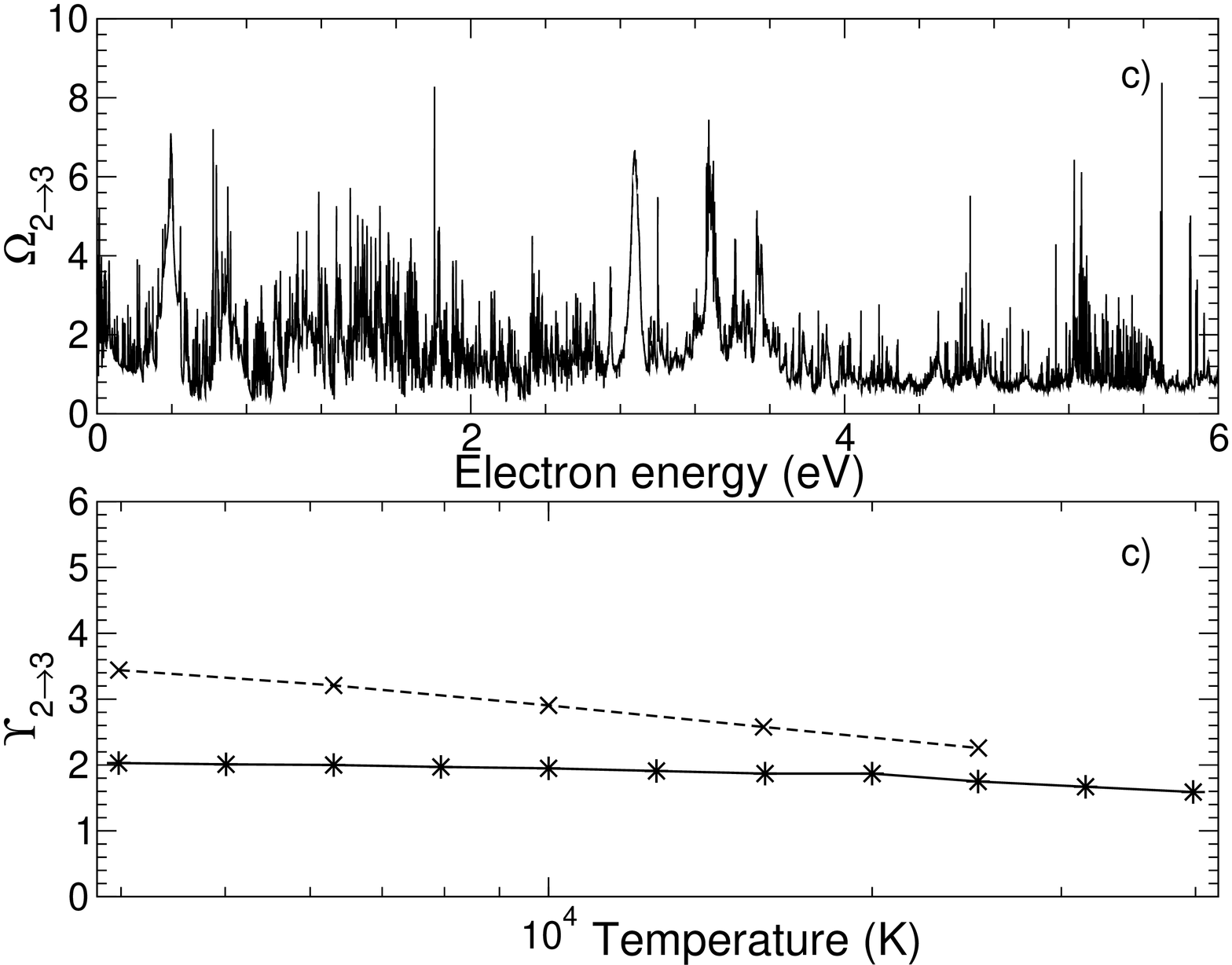}}
        
    \caption{The collision strengths, a) $\Omega_{1\rightarrow 2}$, b) $\Omega_{1\rightarrow 3}$ and c) $\Omega_{2\rightarrow 3}$ presented as a function of electron energy in eV. Also presented are the corresponding effective collision strengths as a function of temperature in K. Solid black lines with stars are the current data set and the dashed line with crosses represent the results from \citet{2016arXiv160200712S}. \label{fig:coll_infra}}
\end{figure}
%

It is necessary to obtain a level of convergence in the spectra of collision strengths by applying a mesh with incremental step sizes in electron energy until convergence is achieved. Initially 5,000 equally spaced energy points were considered and it was found by the time we had reached 40,000 energy points, the $\Upsilon$s had converged for the forbidden transitions among the 292 levels. To extend the energy region we incorporated an additional coarse mesh above the last valence threshold. Due to the long range nature of the Coulomb potential further contributions to the collision strengths arise from the higher partial waves, particularly for the dipole allowed lines. We compute these additional contributions using the \citet{1992A&A...254..436B} sum rule as well as a geomteric series for the long-range non-dipole transitions. Hence converged total collision strengths were accurately generated for all 42,486 transitions among the 292 fine-structure levels included in the collision calculation. The corresponding effective collision strengths were obtained by averaging these finely resolved collision strengths over a Maxwellian distribution of electron velocities for electron temperatures ranging from 3,800 to 40,000 K.  

%
\begin{figure}
\includegraphics[scale=0.31, angle=-90]{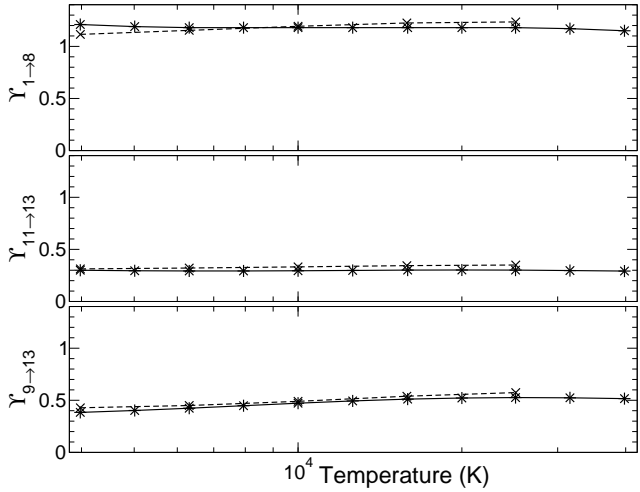}
\caption{The effective collision strengths, $\Upsilon_{1\rightarrow 8}$ (top), $\Upsilon_{11\rightarrow 13}$ (middle) and $\Upsilon_{9\rightarrow 13}$ (bottom), presented as a function of temperature in K (bottom). Solid black lines with stars are the current data set and the dashed line with crosses represent the results from \citep{2016arXiv160200712S}. \label{fig:coll_5to7}}
\end{figure}
%

We present in Figure \ref{fig:coll_infra} the resulting collision strengths and effective collision strengths for a selection of the forbidden near infrared transitions involving the fine-structure split levels of the $^4$F ground state of Co {\sc iii}. Panel (a) represents transition from level 1$\rightarrow$ 2, panel (b) $1 \rightarrow 3$,  and panel (c) $2 \rightarrow 3$.  The collision strength is strongest here for the $\Omega_{1 \rightarrow 2}$ transition due to the $\Delta J = 0$ partial wave. 
Comparison of the effective collision strengths are made for each transition with the work of \citet{2016arXiv160200712S} and good agreement is found for all temperatures above 10,000 K. For temperatures below this value the two calculations deviate somewhat, most probably due to the differing resonance profiles converging onto differing threshold positions. The present calculation has shifted the thresholds to lie in their exact observed positions listed in Table 1. In Figure 6 we present the effective collision strengths for some higher lying transitions, the 3d$^7$ $^4$F$_{9/2}$ - 3d$^7$ $^2$G$_{9/2}$ (1 $\rightarrow$ 8), 3d$^7$ $^2$P$_{1/2}$ - 3d$^7$ $^2$H$_{9/2}$ (11 $\rightarrow$ 13) and 3d$^7$ $^2$G$_{7/2}$ - 3d$^7$ $^2$H$_{9/2}$ (9 $\rightarrow$ 13). Excellent agreement is evident for all three transitions when a comparison is made with the data of \citet{2016arXiv160200712S} across all temperatures where a comparison is possible. The collision strengths and effective collision strengths for all 42,486 forbidden and allowed lines are available from the authors on request but we present in Table \ref{tab:ecollstrengths} the effective collision strengths for all transitions among the lowest 11 3d$^7$ fine-structure levels across nine temperatures of astrophysical importance.

\subsection{Line ratios}
Combining the electron-impact excitation rates with the decay rates ($A$-values), it is possible to investigate important infrared and visible line ratios. For simplicity and purpose of this study, we neglect the recombination and reverse ionization process from the neighbouring ion stages and focus on the populating mechanisms solely of Co {\sc iii}. Assuming local thermal equilibrium and detailed balance, we can derive the collisional radiative modelling approach detailed in \citep{0953-4075-30-15-023}. We can therefore investigate select transitions that have been suggested as temperature or density diagnostics. We define the line ratio as
\[
\mathcal{R} = \frac{I_{j\rightarrow i}}{I_{k\rightarrow m}} = \frac{N_j A_{j\rightarrow i}}{N_kA_{k\rightarrow m}}\frac{\lambda_{mk}}{\lambda_{ij}},
\]
where the populations of the $n^{th}$ level in terms of the ground state are given by $N_n$ and the decay rates are the results from Section 2.1.

\begin{figure}
\centering
\includegraphics[scale=0.31, angle=-90]{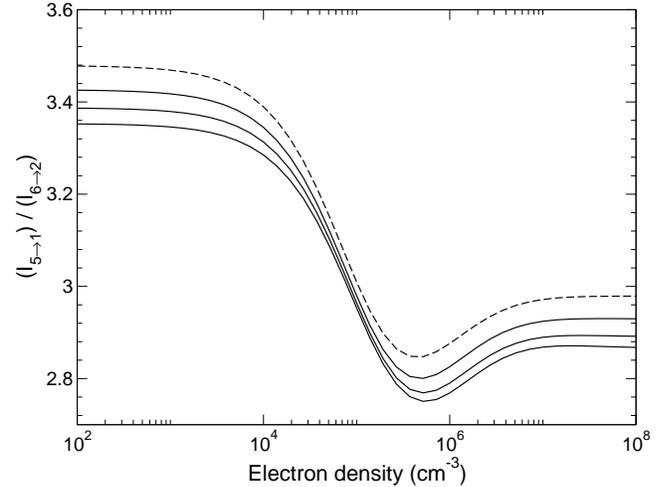}
\caption{The line ratio [0.66$\mu$m]/[0.69$\mu$m] as a function of electron density (cm$^{-3}$). The dashed curve is the lowest temperature, $T = 3,980$K. The remaining solid black curves are for temperatures $T = 5,010$K, $T = 6,310$K, and $T = 7,940$K for decreasing line ratio. \label{fig:consttemp}}
\end{figure}

Figure \ref{fig:consttemp} depicts the results of the line ratio [0.66$\mu$m]/[0.69$\mu$m] which corresponds to the transitions,
\[
\frac{I_{5\rightarrow 1}}{I_{6\rightarrow 2}} = \frac{3\rm{d}^7 ~\rm{a}^4\rm{P}_{5/2}\rightarrow 3\rm{d}^7 ~\rm{a}^4\rm{F}_{9/2}}{3\rm{d}^7 ~\rm{a}^4\rm{P}_{3/2} \rightarrow 3\rm{d}^7 ~\rm{a}^4\rm{F}_{7/2}},
\]
as a function of electron density for particular temperatures. The dashed line provides the lowest temperature, $T = 3,980$K, with increasing $T$ to $T = 5,010$K, $T = 6,310$K and $T = 7,940$K. The line ratio is constant for densities $N_e < 10^{4}$ cm$^{-3}$ and also $N_e > 10^{7}$ cm$^{-3}$ for the temperatures considered. This could be considered a useful diagnostic for densities in the range of $10^{4}$ cm$^{-3}$ $< N_e < 3\times10^{5}$ cm$^{-3}$ where it remains constant for increasing temperatures before it reaches its minimum $\approx 5\times 10^{5}$ cm$^{-3}$

%
%

Next we consider lines suggested by \citet{1995ASPC...78..291R}, (around the wavelength region 6,000 - 6,500 {\rm \AA}) to be applied as density diagnostics. At 10,000 K, we can see the photon emissivity coefficients of the $8\rightarrow 1$ and $8\rightarrow 2$ lines becomes more dominant for increasing temperatures, and are in the ideal wavelength region. \citet{2015ApJ...798...93T} also suggests the $2\rightarrow 1$, 11.88$\mu$m as a useful diagnostic line. We finally present the results of three line ratios ,
\[
\frac{I_{2\rightarrow 1}[11.88\mu m]}{I_{5\rightarrow 1}[0.66\mu m]} = \frac{3\rm{d}^7 ~\rm{a}^4\rm{F}_{7/2} \rightarrow 3\rm{d}^7 ~\rm{a}^4\rm{F}_{9/2}}{3\rm{d}^7 ~\rm{a}^4\rm{P}_{5/2} \rightarrow 3\rm{d}^7 ~\rm{a}^4\rm{F}_{9/2}},
\]
\[
\frac{I_{2\rightarrow 1}[11.88\mu m]}{I_{5\rightarrow 1}[0.59\mu m]} = \frac{3\rm{d}^7 ~\rm{a}^4\rm{F}_{7/2} \rightarrow 3\rm{d}^7 ~\rm{a}^4\rm{F}_{9/2}}{3\rm{d}^7 ~\rm{a}^2\rm{G}_{9/2} \rightarrow 3\rm{d}^7 ~\rm{a}^4\rm{F}_{9/2}},
\]
\[
\frac{I_{2\rightarrow 1}[11.88\mu m]}{I_{5\rightarrow 1}[0.62\mu m]} = \frac{3\rm{d}^7 ~\rm{a}^4\rm{F}_{7/2} \rightarrow 3\rm{d}^7 ~\rm{a}^4\rm{F}_{9/2}}{3\rm{d}^7 ~\rm{a}^2\rm{G}_{9/2} \rightarrow 3\rm{d}^7 ~\rm{a}^4\rm{F}_{7/2}},
\]

%
\begin{figure}
~\\
~\\
  \subfloat{\includegraphics[scale=0.31, angle=-90]{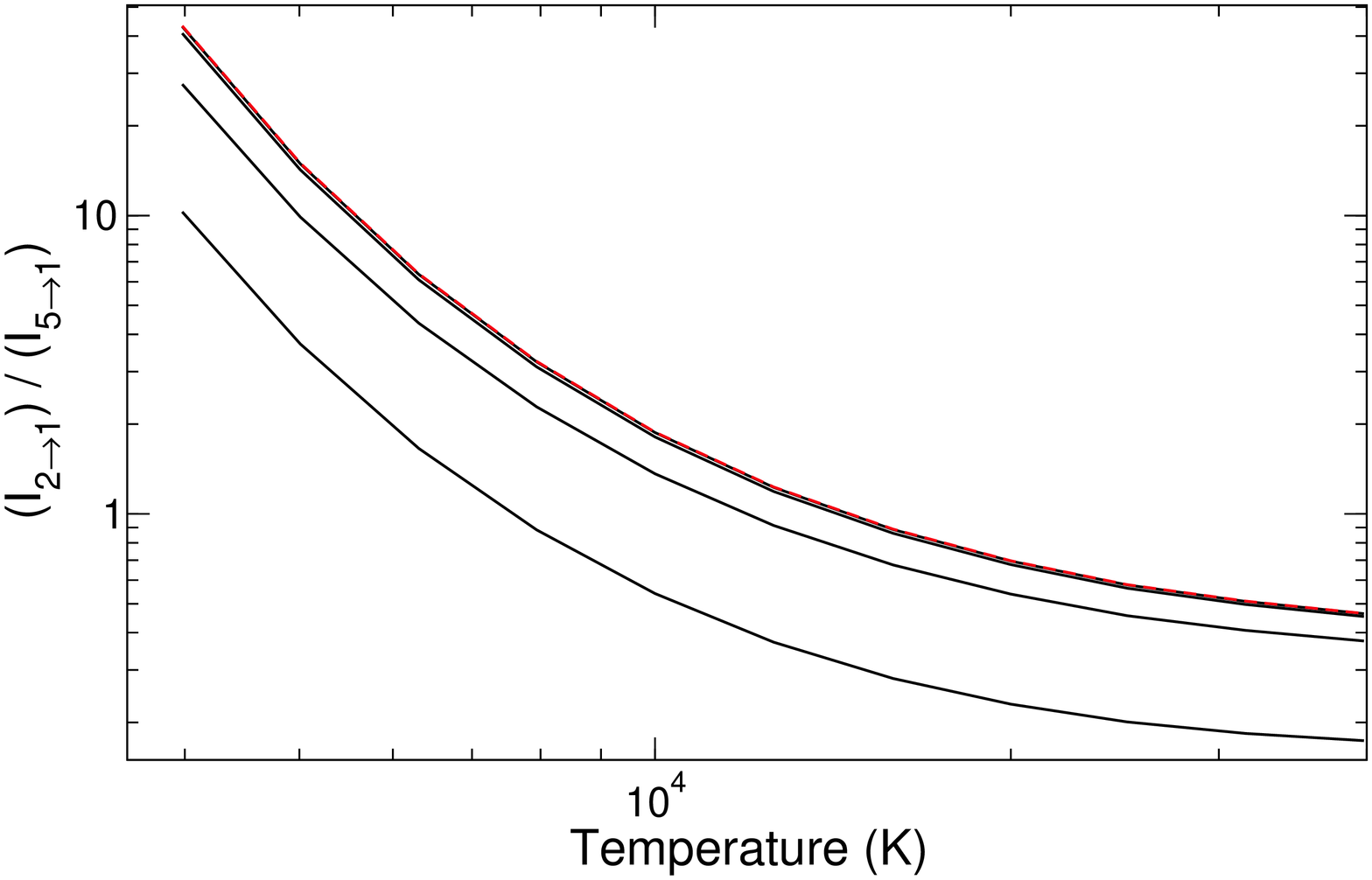}}\\
     \subfloat{\includegraphics[scale=0.31, angle=-90]{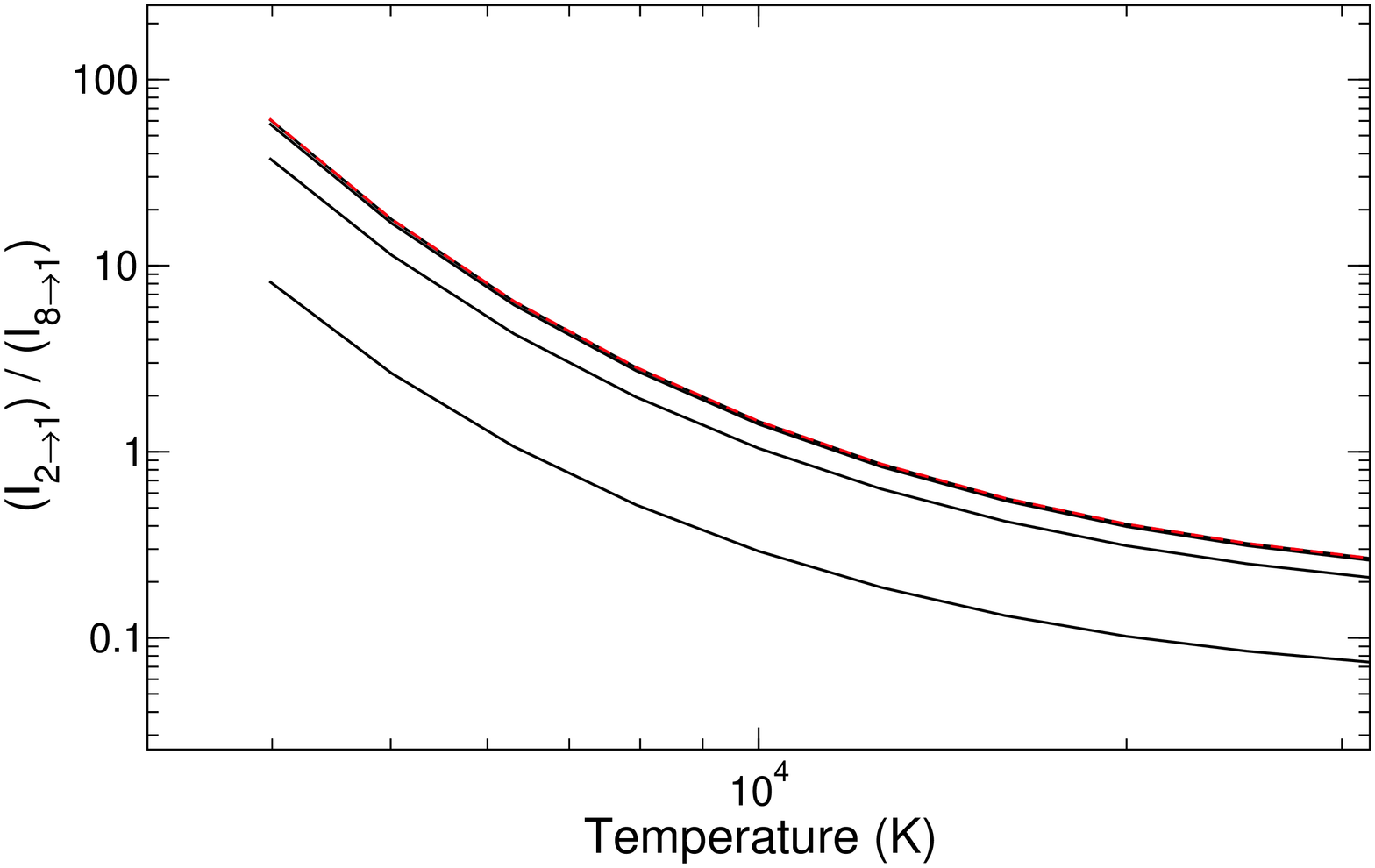}}\\
        \subfloat{\includegraphics[scale=0.31, angle=-90]{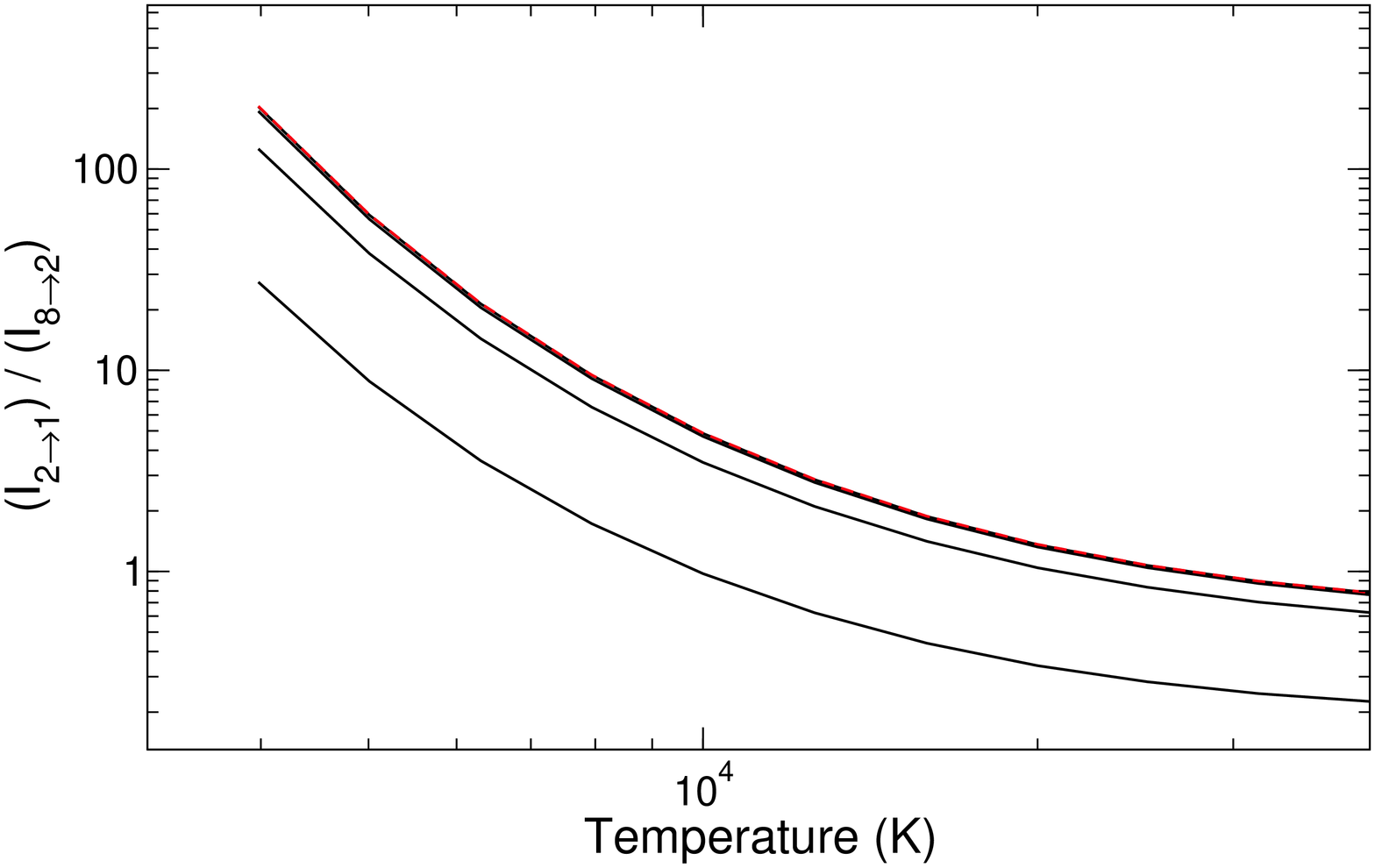}}
        
    \caption{The line ratio [11.88$\mu$m]/[0.66$\mu$m] (top), [11.88$\mu$m]/[0.59$\mu$m] (middle), and [11.88$\mu$m]/[0.62$\mu$m] (bottom) as a function of electron temperature (K). The red, dashed curve is the lowest density, $N_e = 10^2$ cm$^{-3}$. The remaining solid black curves are for temperatures $N_e = 10^3$ cm$^{-3}$, $N_e = 10^4$ cm$^{-3}$, $N_e = 10^5$ cm$^{-3}$, and $N_e = 10^6$ cm$^{-3}$ for decreasing line ratio. \label{fig:constdens}}
\end{figure}
%

We plot Figure \ref{fig:constdens} as a function of electron temperature for all three line ratios. The red dashed curve is the lowest electron density in each subfigure, $N_e = 10^2$ cm$^{-3}$, and the remaining solid curves correspond to $N_e = 10^3$ cm$^{-3}$, $N_e = 10^4$ cm$^{-3}$, $N_e = 10^5$ cm$^{-3}$, and $N_e = 10^6$ cm$^{-3}$. Each line ratio is a varying function of increasing temperature across the range of interest. The line ratios for three lowest densities are generally unchanged, and therefore they provide useful diagnostics for this density range. As $N_e = 10^5$ cm$^{-3}$ and beyond, the line ratio begins to diverge, and becomes constant for even larger densities.

%
%

\section{Conclusion}
In this paper we present an extensive set of atomic data for the photoionization of Co {\sc ii} and the electron-impact excitation of Co {\sc iii}. Initially we have exploited the computer code {\sc grasp0} to obtain a description of the atomic wavefunctions and generate energy levels for the 292 fine-structure bound target states and the corresponding $A$-values for transitions between these levels. Furthermore, the {\sc darc} computer package has been employed to extend the problem to include photon and electron interactions. We present statistically weighted, level-resolved ground and metastable photoionization cross-sections for the Co {\sc ii} ion as well as collision strengths and Maxwellian averaged effective collision strengths describing the electron-impact excitation of the Co {\sc iii} ion. Comparisons are made with other works where possible and good agreement is found where a comparison is available. The reliability of the atomic data presented has been rigorously tested through a variety of means, such as the sophistication of the current calculations where great care has been taken to ensure the inclusion of important correlation and configuration-interaction in the wavefunction expansions. In addition the complex resonance structures in the cross sections (photoionization and excitation) have been accurately resolved through a series of calculations incorporating mesh sizes with finer and finer energy increments. A proper consideration was also taken of the contributions from the high partial waves to ensure convergence of the collision strengths for the allowed transitions in particular (which are not presented in this work). A conclusive assessment of the accuracy of the presented Maxwellian averaged collision strengths will necessarily come from any subsequent astrophysical or diagnostic application. In Section 3.3 the electron-impact excitation rates were combined with the decay rates ($A$-values) to investigate important infrared and visible line ratios. During this process we were able to identify useful transitions that could be used as temperature and/or density sensitive diagnostic lines.

\section*{Acknowledgments}
The work conducted has been supported by STFC through the grant ST/K000802/1. The authors would also like to acknowledge time on the Stuttgart, Hazelhen machine under the PAMOP (44009) project where all calculations have been performed.

\bibliographystyle{mn2e}
\bibliography{mybib.bib}

\begin{thebibliography}{}

\bibitem[\protect\citeauthoryear{{Aggarwal}, {Bogdanovich}, {Karpu{\v
  s}kien{\.e}}, {Keenan}, {Kisielius} \& {Stancalie}}{{Aggarwal}
  et~al.}{2015}]{2015arXiv150907648A}
{Aggarwal} K.~M.,  {Bogdanovich} P.,  {Karpu{\v s}kien{\.e}} R.,  {Keenan}
  F.~P.,  {Kisielius} R.,    {Stancalie} V.,  2015, ArXiv e-prints

\bibitem[\protect\citeauthoryear{{Badnell}}{{Badnell}}{1986}]{1986JPhB...19.3827B}
{Badnell} N.~R.,  1986, J. Phys. B: At. Mol. Phys., 19, 3827

\bibitem[\protect\citeauthoryear{{Bautista}, {Fivet}, {Ballance}, {Quinet},
  {Ferland}, {Mendoza} \& {Kallman}}{{Bautista}
  et~al.}{2015}]{2015ApJ...808..174B}
{Bautista} M.~A.,  {Fivet} V.,  {Ballance} C.,  {Quinet} P.,  {Ferland} G.,
  {Mendoza} C.,    {Kallman} T.~R.,  2015, \apj, 808, 174

\bibitem[\protect\citeauthoryear{{Bergemann}, {Pickering} \&
  {Gehren}}{{Bergemann} et~al.}{2010}]{2010MNRAS.401.1334B}
{Bergemann} M.,  {Pickering} J.~C.,    {Gehren} T.,  2010, \mnras, 401, 1334

\bibitem[\protect\citeauthoryear{{Burgess} \& {Tully}}{{Burgess} \&
  {Tully}}{1992}]{1992A&A...254..436B}
{Burgess} A.,  {Tully} J.~A.,  1992, \aap, 254, 436

\bibitem[\protect\citeauthoryear{{Burke}}{{Burke}}{2011}]{2011rmta.book.....B}
{Burke} P.~G.,  2011, {R-Matrix Theory of Atomic Collisions}.
Springer-Verlag: Berlin Heidelberg

\bibitem[\protect\citeauthoryear{{Burke}, {Hibbert} \& {Robb}}{{Burke}
  et~al.}{1971}]{1971JPhB....4..153B}
{Burke} P.~G.,  {Hibbert} A.,    {Robb} W.~D.,  1971, J. Phys. B: At. Mol.
  Phys., 4, 153

\bibitem[\protect\citeauthoryear{{Burke} \& {Taylor}}{{Burke} \&
  {Taylor}}{1975}]{1975JPhB....8.2620B}
{Burke} P.~G.,  {Taylor} K.~T.,  1975, J. Phys. B: At. Mol. Phys., 8, 2620

\bibitem[\protect\citeauthoryear{{Cardon}, {Smith}, {Scalo}, {Testerman} \&
  {Whaling}}{{Cardon} et~al.}{1982}]{1982ApJ...260..395C}
{Cardon} B.~L.,  {Smith} P.~L.,  {Scalo} J.~M.,  {Testerman} L.,    {Whaling}
  W.,  1982, \apj, 260, 395

\bibitem[\protect\citeauthoryear{{Cowan}}{{Cowan}}{1981}]{1981tass.book.....C}
{Cowan} R.~D.,  1981, {The theory of atomic structure and spectra}.
Berkeley: Univ. California Press

\bibitem[\protect\citeauthoryear{{Dessart}, {Hillier}, {Blondin} \&
  {Khokhlov}}{{Dessart} et~al.}{2014}]{2014MNRAS.441.3249D}
{Dessart} L.,  {Hillier} D.~J.,  {Blondin} S.,    {Khokhlov} A.,  2014, \mnras,
  441, 3249

\bibitem[\protect\citeauthoryear{{Diamond}, {Hoeflich} \& {Gerardy}}{{Diamond}
  et~al.}{2015}]{2015ApJ...806..107D}
{Diamond} T.~R.,  {Hoeflich} P.,    {Gerardy} C.~L.,  2015, \apj, 806, 107

\bibitem[\protect\citeauthoryear{{Dworetsky}}{{Dworetsky}}{1982}]{1982Obs...102..138D}
{Dworetsky} M.~M.,  1982, Observatory, 102, 138

\bibitem[\protect\citeauthoryear{{Dworetsky}, {Trueman} \&
  {Stickland}}{{Dworetsky} et~al.}{1980}]{1980A&A....85..138D}
{Dworetsky} M.~M.,  {Trueman} M.~R.~G.,    {Stickland} D.~J.,  1980, \aap, 85,
  138

\bibitem[\protect\citeauthoryear{{Eissner}, {Jones} \& {Nussbaumer}}{{Eissner}
  et~al.}{1974}]{1974CoPhC...8..270E}
{Eissner} W.,  {Jones} M.,    {Nussbaumer} H.,  1974, Computer Physics
  Communications, 8, 270

\bibitem[\protect\citeauthoryear{{Federman}, {Sheffer}, {Lambert} \&
  {Gilliland}}{{Federman} et~al.}{1993}]{1993ApJ...413L..51F}
{Federman} S.~R.,  {Sheffer} Y.,  {Lambert} D.~L.,    {Gilliland} R.~L.,  1993,
  \apjl, 413, L51

\bibitem[\protect\citeauthoryear{{Fern{\'a}ndez-Menchero}, {Del Zanna} \&
  {Badnell}}{{Fern{\'a}ndez-Menchero} et~al.}{2015}]{2015MNRAS.450.4174F}
{Fern{\'a}ndez-Menchero} L.,  {Del Zanna} G.,    {Badnell} N.~R.,  2015,
  \mnras, 450, 4174

\bibitem[\protect\citeauthoryear{{Fivet}, {Quinet} \& {Bautista}}{{Fivet}
  et~al.}{2016}]{2016A&A...585A.121F}
{Fivet} V.,  {Quinet} P.,    {Bautista} M.~A.,  2016, \aap, 585, A121

\bibitem[\protect\citeauthoryear{{Fontes}, {Zhang}, {Abdallah} Jr., {Clark},
  {Kilcrease}, {Colgan}, {Cunningham}, {Hakel}, {Magee} \& {Sherrill}}{{Fontes}
  et~al.}{2015}]{2015JPhB...48n4014F}
{Fontes} C.~J.,  {Zhang} H.~L.,  {Abdallah} Jr. J.,  {Clark} R.~E.~H.,
  {Kilcrease} D.~P.,  {Colgan} J.,  {Cunningham} R.~T.,  {Hakel} P.,  {Magee}
  N.~H.,    {Sherrill} M.~E.,  2015, J. Phys. B: At. Mol. Phys., 48, 144014

\bibitem[\protect\citeauthoryear{{Gerardy}, {Meikle}, {Kotak}, {H{\"o}flich},
  {Farrah}, {Filippenko}, {Foley}, {Lundqvist}, {Mattila}, {Pozzo},
  {Sollerman}, {Van Dyk} \& {Wheeler}}{{Gerardy}
  et~al.}{2007}]{2007ApJ...661..995G}
{Gerardy} C.~L.,  {Meikle} W.~P.~S.,  {Kotak} R.,  {H{\"o}flich} P.,  {Farrah}
  D.,  {Filippenko} A.~V.,  {Foley} R.~J.,  {Lundqvist} P.,  {Mattila} S.,
  {Pozzo} M.,  {Sollerman} J.,  {Van Dyk} S.~D.,    {Wheeler} J.~C.,  2007,
  \apj, 661, 995

\bibitem[\protect\citeauthoryear{{Gharaibeh}, {Bizau}, {Cubaynes}, {Guilbaud},
  {El Hassan}, {Shorman}, {Miron}, {Nicolas}, {Robert}, {Blancard} \&
  {McLaughlin}}{{Gharaibeh} et~al.}{2011}]{2011JPhB...44q5208G}
{Gharaibeh} M.~F.,  {Bizau} J.~M.,  {Cubaynes} D.,  {Guilbaud} S.,  {El Hassan}
  N.,  {Shorman} M.~M.~A.,  {Miron} C.,  {Nicolas} C.,  {Robert} E.,
  {Blancard} C.,    {McLaughlin} B.~M.,  2011, J. Phys. B: At. Mol. Phys., 44,
  175208

\bibitem[\protect\citeauthoryear{{Griffin}, {Badnell} \& {Pindzola}}{{Griffin}
  et~al.}{1998}]{1998JPhB...31.3713G}
{Griffin} D.~C.,  {Badnell} N.~R.,    {Pindzola} M.~S.,  1998, J. Phys. B: At.
  Mol. Phys., 31, 3713

\bibitem[\protect\citeauthoryear{Griffin, Pindzola, Shaw, Badnell, O'Mullane \&
  Summers}{Griffin et~al.}{1997}]{0953-4075-30-15-023}
Griffin D.~C.,  Pindzola M.~S.,  Shaw J.~A.,  Badnell N.~R.,  O'Mullane M.,
  Summers H.~P.,  1997, J. Phys. B: At. Mol. Phys., 30, 3543

\bibitem[\protect\citeauthoryear{{Hansen}, {Raassen} \& {Uylings}}{{Hansen}
  et~al.}{1984}]{1984ApJ...277..435H}
{Hansen} J.~E.,  {Raassen} A.~J.~J.,    {Uylings} P.~H.~M.,  1984, \apj, 277,
  435

\bibitem[\protect\citeauthoryear{{Hillier}}{{Hillier}}{2011}]{2011Ap&SS.336...87H}
{Hillier} D.~J.,  2011, \apss, 336, 87

\bibitem[\protect\citeauthoryear{{Li}, {McCray} \& {Sunyaev}}{{Li}
  et~al.}{1993}]{1993ApJ...419..824L}
{Li} H.,  {McCray} R.,    {Sunyaev} R.~A.,  1993, \apj, 419, 824

\bibitem[\protect\citeauthoryear{{Mazzali}, {Chugai}, {Turatto}, {Lucy},
  {Danziger}, {Cappellaro}, {della Valle} \& {Benetti}}{{Mazzali}
  et~al.}{1997}]{1997MNRAS.284..151M}
{Mazzali} P.~A.,  {Chugai} N.,  {Turatto} M.,  {Lucy} L.~B.,  {Danziger} I.~J.,
   {Cappellaro} E.,  {della Valle} M.,    {Benetti} S.,  1997, \mnras, 284, 151

\bibitem[\protect\citeauthoryear{{Meikle}, {Spyromilio}, {Varani} \&
  {Allen}}{{Meikle} et~al.}{1989}]{1989MNRAS.238..193M}
{Meikle} W.~P.~S.,  {Spyromilio} J.,  {Varani} G.~F.,    {Allen} D.~A.,  1989,
  \mnras, 238, 193

\bibitem[\protect\citeauthoryear{{M{\"u}ller}, {Schippers}, {Phaneuf},
  {Scully}, {Aguilar}, {Covington}, {{\'A}lvarez}, {Cisneros}, {Emmons},
  {Gharaibeh}, {Hinojosa}, {Schlachter} \& {McLaughlin}}{{M{\"u}ller}
  et~al.}{2009}]{2009JPhB...42w5602M}
{M{\"u}ller} A.,  {Schippers} S.,  {Phaneuf} R.~A.,  {Scully} S.~W.~J.,
  {Aguilar} A.,  {Covington} A.~M.,  {{\'A}lvarez} I.,  {Cisneros} C.,
  {Emmons} E.~D.,  {Gharaibeh} M.~F.,  {Hinojosa} G.,  {Schlachter} A.~S.,
  {McLaughlin} B.~M.,  2009, J. Phys. B: At. Mol. Phys., 42, 235602

\bibitem[\protect\citeauthoryear{{Nussbaumer} \& {Storey}}{{Nussbaumer} \&
  {Storey}}{1988a}]{1988A&A...193..327N}
{Nussbaumer} H.,  {Storey} P.~J.,  1988a, \aap, 193, 327

\bibitem[\protect\citeauthoryear{{Nussbaumer} \& {Storey}}{{Nussbaumer} \&
  {Storey}}{1988b}]{1988A&A...200L..25N}
{Nussbaumer} H.,  {Storey} P.~J.,  1988b, \aap, 200, L25

\bibitem[\protect\citeauthoryear{{Pickering}, {Raassen}, {Uylings} \&
  {Johansson}}{{Pickering} et~al.}{1998}]{1998ApJS..117..261P}
{Pickering} J.~C.,  {Raassen} A.~J.~J.,  {Uylings} P.~H.~M.,    {Johansson} S.,
   1998, \apjs, 117, 261

\bibitem[\protect\citeauthoryear{{Quinet}}{{Quinet}}{1998}]{1998A&AS..129..147Q}
{Quinet} P.,  1998, \aaps, 129, 147

\bibitem[\protect\citeauthoryear{{Ramsbottom}}{{Ramsbottom}}{2009}]{2009ADNDT..95..910R}
{Ramsbottom} C.~A.,  2009, ADNDT, 95, 910

\bibitem[\protect\citeauthoryear{{Ramsbottom}, {Hudson}, {Norrington} \&
  {Scott}}{{Ramsbottom} et~al.}{2007}]{2007A&A...475..765R}
{Ramsbottom} C.~A.,  {Hudson} C.~E.,  {Norrington} P.~H.,    {Scott} M.~P.,
  2007, \aap, 475, 765

\bibitem[\protect\citeauthoryear{{Reilman} \& {Manson}}{{Reilman} \&
  {Manson}}{1979}]{1979ApJS...40..815R}
{Reilman} R.~F.,  {Manson} S.~T.,  1979, \apjs, 40, 815

\bibitem[\protect\citeauthoryear{{Ruiz-Lapuente}}{{Ruiz-Lapuente}}{1995}]{1995ASPC...78..291R}
{Ruiz-Lapuente} P.,  1995, in Book title here Vol.~78 of ASP Conf. Ser.,
  {Calculated Late-time Spectra of Type IA Supernovae: Atomic Needs and
  Astrophysical Interest}.
p.~291

\bibitem[\protect\citeauthoryear{{Scott} \& {Burke}}{{Scott} \&
  {Burke}}{1980}]{1980JPhB...13.4299S}
{Scott} N.~S.,  {Burke} P.~G.,  1980, J. Phys. B: At. Mol. Phys., 13, 4299

\bibitem[\protect\citeauthoryear{{Smith} \& {Dworetsky}}{{Smith} \&
  {Dworetsky}}{1993}]{1993A&A...274..335S}
{Smith} K.~C.,  {Dworetsky} M.~M.,  1993, \aap, 274, 335

\bibitem[\protect\citeauthoryear{{Snow} Jr., {Weiler} \& {Oegerle}}{{Snow}
  et~al.}{1979}]{1979ApJ...234..506S}
{Snow} Jr. T.~P.,  {Weiler} E.~J.,    {Oegerle} W.~R.,  1979, \apj, 234, 506

\bibitem[\protect\citeauthoryear{{Storey} \& {Sochi}}{{Storey} \&
  {Sochi}}{2016}]{2016arXiv160200712S}
{Storey} P.~J.,  {Sochi} T.,  2016, ArXiv e-prints

\bibitem[\protect\citeauthoryear{{Storey}, {Zeippen} \& {Sochi}}{{Storey}
  et~al.}{2015}]{2015arXiv150903164S}
{Storey} P.~J.,  {Zeippen} C.~J.,    {Sochi} T.,  2015, ArXiv e-prints

\bibitem[\protect\citeauthoryear{{Sugar} \& {Corliss}}{{Sugar} \&
  {Corliss}}{1985}]{1985aeli.book.....S}
{Sugar} J.,  {Corliss} C.,  1985, {Atomic energy levels of the iron-period
  elements: Potassium through Nickel}.
Washington: American Chemical Soc.

\bibitem[\protect\citeauthoryear{{Telesco}, {H{\"o}flich}, {Li}, {{\'A}lvarez},
  {Wright}, {Barnes}, {Fern{\'a}ndez}, {Hough}, {Levenson}, {Mari{\~n}as},
  {Packham}, {Pantin}, {Rebolo}, {Roche} \& {Zhang}}{{Telesco}
  et~al.}{2015}]{2015ApJ...798...93T}
{Telesco} C.~M.,  {H{\"o}flich} P.,  {Li} D.,  {{\'A}lvarez} C.,  {Wright}
  C.~M.,  {Barnes} P.~J.,  {Fern{\'a}ndez} S.,  {Hough} J.~H.,  {Levenson}
  N.~A.,  {Mari{\~n}as} N.,  {Packham} C.,  {Pantin} E.,  {Rebolo} R.,  {Roche}
  P.,    {Zhang} H.,  2015, \apj, 798, 93

\bibitem[\protect\citeauthoryear{{Thackeray}}{{Thackeray}}{1976}]{1976MNRAS.174P..59T}
{Thackeray} A.~D.,  1976, \mnras, 174, 59P

\bibitem[\protect\citeauthoryear{{Tyndall}, {Ramsbottom}, {Ballance} \&
  {Hibbert}}{{Tyndall} et~al.}{2016}]{2016MNRAS.456..366T}
{Tyndall} N.~B.,  {Ramsbottom} C.~A.,  {Ballance} C.~P.,    {Hibbert} A.,
  2016, \mnras, 456, 366

\bibitem[\protect\citeauthoryear{{Verner}, {Yakovlev}, {Band} \&
  {Trzhaskovskaya}}{{Verner} et~al.}{1993}]{1993ADNDT..55..233V}
{Verner} D.~A.,  {Yakovlev} D.~G.,  {Band} I.~M.,    {Trzhaskovskaya} M.~B.,
  1993, ADNDT, 55, 233

\bibitem[\protect\citeauthoryear{{Zatsarinny} \& {Bartschat}}{{Zatsarinny} \&
  {Bartschat}}{2004}]{2004JPhB...37.4693Z}
{Zatsarinny} O.,  {Bartschat} K.,  2004, J. Phys. B: At. Mol. Phys., 37, 4693

\bibitem[\protect\citeauthoryear{{Zatsarinny} \& {Froese Fischer}}{{Zatsarinny}
  \& {Froese Fischer}}{2000}]{2000JPhB...33..313Z}
{Zatsarinny} O.,  {Froese Fischer} C.,  2000, J. Phys. B: At. Mol. Phys., 33,
  313

\bibitem[\protect\citeauthoryear{{Zethson}, {Gull}, {Hartman}, {Johansson},
  {Davidson} \& {Ishibashi}}{{Zethson} et~al.}{2001}]{2001AJ....122..322Z}
{Zethson} T.,  {Gull} T.~R.,  {Hartman} H.,  {Johansson} S.,  {Davidson} K.,
  {Ishibashi} K.,  2001, \aj, 122, 322

\end{thebibliography}

\newpage
\begin{table*}
\centering
\begin{tabular}{c c c c c c c c c c c}
\toprule
\multicolumn{2}{c}{ } & \multicolumn{9}{c}{log $T$ (K)} \\
$i$ & $j$   & 3.6 & 3.7 & 3.8 & 3.9 &  4 & 4.1 & 4.2 & 4.3 & 4.4 \\
\midrule
   1 &    2 & $ 2.92^{+00}$ & $ 2.90^{+00}$ & $ 2.86^{+00}$ & $ 2.82^{+00}$ & $ 2.77^{+00}$ & $ 2.70^{+00}$ & $ 2.63^{+00}$ & $ 2.54^{+00}$ & $ 2.44^{+00}$\\
   1 &    3 & $ 9.24^{-01}$ & $ 9.36^{-01}$ & $ 9.41^{-01}$ & $ 9.37^{-01}$ & $ 9.26^{-01}$ & $ 9.06^{-01}$ & $ 8.79^{-01}$ & $ 8.45^{-01}$ & $ 8.06^{-01}$\\
   2 &    3 & $ 2.02^{+00}$ & $ 2.01^{+00}$ & $ 1.99^{+00}$ & $ 1.97^{+00}$ & $ 1.94^{+00}$ & $ 1.91^{+00}$ & $ 1.86^{+00}$ & $ 1.81^{+00}$ & $ 1.74^{+00}$\\ 
   1 &    4 & $ 3.17^{-01}$ & $ 3.23^{-01}$ & $ 3.26^{-01}$ & $ 3.25^{-01}$ & $ 3.21^{-01}$ & $ 3.13^{-01}$ & $ 3.03^{-01}$ & $ 2.90^{-01}$ & $ 2.75^{-01}$\\
   2 &    4 & $ 7.46^{-01}$ & $ 7.65^{-01}$ & $ 7.77^{-01}$ & $ 7.80^{-01}$ & $ 7.76^{-01}$ & $ 7.64^{-01}$ & $ 7.43^{-01}$ & $ 7.17^{-01}$ & $ 6.85^{-01}$\\
   3 &    4 & $ 1.42^{+00}$ & $ 1.42^{+00}$ & $ 1.42^{+00}$ & $ 1.42^{+00}$ & $ 1.40^{+00}$ & $ 1.39^{+00}$ & $ 1.36^{+00}$ & $ 1.32^{+00}$ & $ 1.28^{+00}$\\
   1 &    5 & $ 1.16^{+00}$ & $ 1.16^{+00}$ & $ 1.17^{+00}$ & $ 1.19^{+00}$ & $ 1.20^{+00}$ & $ 1.21^{+00}$ & $ 1.21^{+00}$ & $ 1.20^{+00}$ & $ 1.19^{+00}$\\
   2 &    5 & $ 8.11^{-01}$ & $ 8.03^{-01}$ & $ 8.02^{-01}$ & $ 8.07^{-01}$ & $ 8.13^{-01}$ & $ 8.17^{-01}$ & $ 8.15^{-01}$ & $ 8.05^{-01}$ & $ 7.89^{-01}$\\
   3 &    5 & $ 5.72^{-01}$ & $ 5.60^{-01}$ & $ 5.53^{-01}$ & $ 5.51^{-01}$ & $ 5.50^{-01}$ & $ 5.48^{-01}$ & $ 5.43^{-01}$ & $ 5.33^{-01}$ & $ 5.19^{-01}$\\   
   4 &    5 & $ 3.68^{-01}$ & $ 3.51^{-01}$ & $ 3.37^{-01}$ & $ 3.27^{-01}$ & $ 3.20^{-01}$ & $ 3.13^{-01}$ & $ 3.06^{-01}$ & $ 2.98^{-01}$ & $ 2.88^{-01}$\\    
   1 &    6 & $ 4.84^{-01}$ & $ 4.82^{-01}$ & $ 4.84^{-01}$ & $ 4.90^{-01}$ & $ 4.95^{-01}$ & $ 4.98^{-01}$ & $ 4.97^{-01}$ & $ 4.91^{-01}$ & $ 4.82^{-01}$\\
   2 &    6 & $ 5.54^{-01}$ & $ 5.52^{-01}$ & $ 5.53^{-01}$ & $ 5.56^{-01}$ & $ 5.60^{-01}$ & $ 5.61^{-01}$ & $ 5.58^{-01}$ & $ 5.52^{-01}$ & $ 5.41^{-01}$\\   
   3 &    6 & $ 4.55^{-01}$ & $ 4.51^{-01}$ & $ 4.51^{-01}$ & $ 4.54^{-01}$ & $ 4.57^{-01}$ & $ 4.59^{-01}$ & $ 4.57^{-01}$ & $ 4.52^{-01}$ & $ 4.44^{-01}$\\   
   4 &    6 & $ 2.92^{-01}$ & $ 2.88^{-01}$ & $ 2.89^{-01}$ & $ 2.92^{-01}$ & $ 2.96^{-01}$ & $ 3.00^{-01}$ & $ 3.02^{-01}$ & $ 3.00^{-01}$ & $ 2.96^{-01}$\\
   5 &    6 & $ 6.14^{-01}$ & $ 6.15^{-01}$ & $ 6.21^{-01}$ & $ 6.31^{-01}$ & $ 6.44^{-01}$ & $ 6.59^{-01}$ & $ 6.73^{-01}$ & $ 6.82^{-01}$ & $ 6.87^{-01}$\\     
   1 &    7 & $ 1.86^{-01}$ & $ 1.85^{-01}$ & $ 1.85^{-01}$ & $ 1.86^{-01}$ & $ 1.87^{-01}$ & $ 1.86^{-01}$ & $ 1.84^{-01}$ & $ 1.81^{-01}$ & $ 1.76^{-01}$\\
   2 &    7 & $ 2.03^{-01}$ & $ 2.03^{-01}$ & $ 2.06^{-01}$ & $ 2.10^{-01}$ & $ 2.15^{-01}$ & $ 2.18^{-01}$ & $ 2.19^{-01}$ & $ 2.18^{-01}$ & $ 2.14^{-01}$\\
   3 &    7 & $ 2.34^{-01}$ & $ 2.36^{-01}$ & $ 2.39^{-01}$ & $ 2.43^{-01}$ & $ 2.48^{-01}$ & $ 2.51^{-01}$ & $ 2.53^{-01}$ & $ 2.52^{-01}$ & $ 2.48^{-01}$\\      
   4 &    7 & $ 2.19^{-01}$ & $ 2.20^{-01}$ & $ 2.22^{-01}$ & $ 2.26^{-01}$ & $ 2.29^{-01}$ & $ 2.31^{-01}$ & $ 2.32^{-01}$ & $ 2.31^{-01}$ & $ 2.28^{-01}$\\
   5 &    7 & $ 2.43^{-01}$ & $ 2.45^{-01}$ & $ 2.49^{-01}$ & $ 2.56^{-01}$ & $ 2.65^{-01}$ & $ 2.75^{-01}$ & $ 2.85^{-01}$ & $ 2.93^{-01}$ & $ 2.98^{-01}$\\      
   6 &    7 & $ 2.73^{-01}$ & $ 2.72^{-01}$ & $ 2.72^{-01}$ & $ 2.74^{-01}$ & $ 2.76^{-01}$ & $ 2.79^{-01}$ & $ 2.81^{-01}$ & $ 2.83^{-01}$ & $ 2.82^{-01}$\\   
   1 &    8 & $ 1.22^{+00}$ & $ 1.20^{+00}$ & $ 1.19^{+00}$ & $ 1.19^{+00}$ & $ 1.19^{+00}$ & $ 1.19^{+00}$ & $ 1.19^{+00}$ & $ 1.19^{+00}$ & $ 1.18^{+00}$\\
   2 &    8 & $ 7.27^{-01}$ & $ 7.14^{-01}$ & $ 7.06^{-01}$ & $ 7.02^{-01}$ & $ 6.99^{-01}$ & $ 6.98^{-01}$ & $ 6.96^{-01}$ & $ 6.92^{-01}$ & $ 6.86^{-01}$\\   
   3 &    8 & $ 3.85^{-01}$ & $ 3.76^{-01}$ & $ 3.70^{-01}$ & $ 3.65^{-01}$ & $ 3.62^{-01}$ & $ 3.59^{-01}$ & $ 3.55^{-01}$ & $ 3.51^{-01}$ & $ 3.45^{-01}$\\   
   4 &    8 & $ 1.75^{-01}$ & $ 1.69^{-01}$ & $ 1.65^{-01}$ & $ 1.60^{-01}$ & $ 1.57^{-01}$ & $ 1.53^{-01}$ & $ 1.50^{-01}$ & $ 1.46^{-01}$ & $ 1.43^{-01}$\\
   5 &    8 & $ 3.24^{-01}$ & $ 3.18^{-01}$ & $ 3.11^{-01}$ & $ 3.07^{-01}$ & $ 3.04^{-01}$ & $ 3.02^{-01}$ & $ 3.01^{-01}$ & $ 3.00^{-01}$ & $ 2.98^{-01}$\\
   6 &    8 & $ 1.56^{-01}$ & $ 1.51^{-01}$ & $ 1.47^{-01}$ & $ 1.43^{-01}$ & $ 1.41^{-01}$ & $ 1.40^{-01}$ & $ 1.39^{-01}$ & $ 1.38^{-01}$ & $ 1.37^{-01}$\\
   7 &    8 & $ 5.48^{-02}$ & $ 5.20^{-02}$ & $ 4.93^{-02}$ & $ 4.70^{-02}$ & $ 4.51^{-02}$ & $ 4.36^{-02}$ & $ 4.24^{-02}$ & $ 4.14^{-02}$ & $ 4.03^{-02}$\\            
   1 &    9 & $ 3.16^{-01}$ & $ 3.10^{-01}$ & $ 3.06^{-01}$ & $ 3.05^{-01}$ & $ 3.05^{-01}$ & $ 3.04^{-01}$ & $ 3.03^{-01}$ & $ 3.01^{-01}$ & $ 2.97^{-01}$\\
   2 &    9 & $ 5.02^{-01}$ & $ 4.99^{-01}$ & $ 4.99^{-01}$ & $ 5.01^{-01}$ & $ 5.05^{-01}$ & $ 5.08^{-01}$ & $ 5.11^{-01}$ & $ 5.12^{-01}$ & $ 5.10^{-01}$\\   
   3 &    9 & $ 5.33^{-01}$ & $ 5.29^{-01}$ & $ 5.29^{-01}$ & $ 5.31^{-01}$ & $ 5.34^{-01}$ & $ 5.37^{-01}$ & $ 5.40^{-01}$ & $ 5.41^{-01}$ & $ 5.39^{-01}$\\
   4 &    9 & $ 4.29^{-01}$ & $ 4.27^{-01}$ & $ 4.26^{-01}$ & $ 4.27^{-01}$ & $ 4.29^{-01}$ & $ 4.32^{-01}$ & $ 4.33^{-01}$ & $ 4.34^{-01}$ & $ 4.32^{-01}$\\
   5 &    9 & $ 1.53^{-01}$ & $ 1.46^{-01}$ & $ 1.41^{-01}$ & $ 1.37^{-01}$ & $ 1.36^{-01}$ & $ 1.36^{-01}$ & $ 1.36^{-01}$ & $ 1.37^{-01}$ & $ 1.36^{-01}$\\
   6 &    9 & $ 1.72^{-01}$ & $ 1.66^{-01}$ & $ 1.62^{-01}$ & $ 1.58^{-01}$ & $ 1.56^{-01}$ & $ 1.56^{-01}$ & $ 1.55^{-01}$ & $ 1.55^{-01}$ & $ 1.55^{-01}$\\
   7 &    9 & $ 1.09^{-01}$ & $ 1.06^{-01}$ & $ 1.04^{-01}$ & $ 1.02^{-01}$ & $ 1.01^{-01}$ & $ 9.99^{-02}$ & $ 9.95^{-02}$ & $ 9.92^{-02}$ & $ 9.86^{-02}$\\
      8 &    9 & $ 1.29^{+00}$ & $ 1.27^{+00}$ & $ 1.26^{+00}$ & $ 1.26^{+00}$ & $ 1.27^{+00}$ & $ 1.28^{+00}$ & $ 1.29^{+00}$ & $ 1.29^{+00}$ & $ 1.28^{+00}$\\                  
  1 &  10 & $ 3.06^{-01}$ & $ 3.04^{-01}$ & $ 3.04^{-01}$ & $ 3.03^{-01}$ & $ 3.03^{-01}$ & $ 3.02^{-01}$ & $ 3.00^{-01}$ & $ 2.99^{-01}$ & $ 2.97^{-01}$\\
   2 &   10 & $ 2.76^{-01}$ & $ 2.79^{-01}$ & $ 2.82^{-01}$ & $ 2.86^{-01}$ & $ 2.88^{-01}$ & $ 2.89^{-01}$ & $ 2.87^{-01}$ & $ 2.83^{-01}$ & $ 2.78^{-01}$\\
   3 &   10 & $ 2.05^{-01}$ & $ 2.09^{-01}$ & $ 2.13^{-01}$ & $ 2.18^{-01}$ & $ 2.21^{-01}$ & $ 2.21^{-01}$ & $ 2.20^{-01}$ & $ 2.16^{-01}$ & $ 2.10^{-01}$\\   
   4 &   10 & $ 1.29^{-01}$ & $ 1.32^{-01}$ & $ 1.36^{-01}$ & $ 1.40^{-01}$ & $ 1.43^{-01}$ & $ 1.43^{-01}$ & $ 1.42^{-01}$ & $ 1.39^{-01}$ & $ 1.34^{-01}$\\
   5 &   10 & $ 2.85^{-01}$ & $ 2.84^{-01}$ & $ 2.86^{-01}$ & $ 2.89^{-01}$ & $ 2.92^{-01}$ & $ 2.95^{-01}$ & $ 2.95^{-01}$ & $ 2.93^{-01}$ & $ 2.89^{-01}$\\
   6 &   10 & $ 2.29^{-01}$ & $ 2.27^{-01}$ & $ 2.27^{-01}$ & $ 2.29^{-01}$ & $ 2.31^{-01}$ & $ 2.32^{-01}$ & $ 2.31^{-01}$ & $ 2.29^{-01}$ & $ 2.25^{-01}$\\
   7  &   10 & $ 9.14^{-02}$ & $ 9.23^{-02}$ & $ 9.39^{-02}$ & $ 9.56^{-02}$ & $ 9.69^{-02}$ & $ 9.75^{-02}$ & $ 9.72^{-02}$ & $ 9.60^{-02}$ & $ 9.40^{-02}$\\
   8 &   10 & $ 4.12^{-01}$ & $ 4.06^{-01}$ & $ 4.01^{-01}$ & $ 3.99^{-01}$ & $ 3.99^{-01}$ & $ 4.00^{-01}$ & $ 4.03^{-01}$ & $ 4.07^{-01}$ & $ 4.09^{-01}$\\
   9 &   10 & $ 3.20^{-01}$ & $ 3.17^{-01}$ & $ 3.17^{-01}$ & $ 3.20^{-01}$ & $ 3.26^{-01}$ & $ 3.34^{-01}$ & $ 3.42^{-01}$ & $ 3.47^{-01}$ & $ 3.50^{-01}$\\                    
  1 &   11 & $ 8.09^{-02}$ & $ 8.03^{-02}$ & $ 8.01^{-02}$ & $ 8.02^{-02}$ & $ 8.05^{-02}$ & $ 8.08^{-02}$ & $ 8.09^{-02}$ & $ 8.08^{-02}$ & $ 8.03^{-02}$\\
   2 &   11 & $ 1.09^{-01}$ & $ 1.10^{-01}$ & $ 1.11^{-01}$ & $ 1.11^{-01}$ & $ 1.12^{-01}$ & $ 1.12^{-01}$ & $ 1.12^{-01}$ & $ 1.11^{-01}$ & $ 1.09^{-01}$\\
   3 &   11 & $ 1.16^{-01}$ & $ 1.19^{-01}$ & $ 1.22^{-01}$ & $ 1.26^{-01}$ & $ 1.28^{-01}$ & $ 1.29^{-01}$ & $ 1.28^{-01}$ & $ 1.27^{-01}$ & $ 1.24^{-01}$\\
   4 &   11 & $ 9.23^{-02}$ & $ 9.71^{-02}$ & $ 1.02^{-01}$ & $ 1.07^{-01}$ & $ 1.11^{-01}$ & $ 1.13^{-01}$ & $ 1.13^{-01}$ & $ 1.11^{-01}$ & $ 1.08^{-01}$\\
   5 &   11 & $ 7.49^{-02}$ & $ 7.54^{-02}$ & $ 7.66^{-02}$ & $ 7.83^{-02}$ & $ 8.00^{-02}$ & $ 8.14^{-02}$ & $ 8.23^{-02}$ & $ 8.23^{-02}$ & $ 8.14^{-02}$\\
   6 &   11 & $ 8.77^{-02}$ & $ 8.98^{-02}$ & $ 9.27^{-02}$ & $ 9.61^{-02}$ & $ 9.93^{-02}$ & $ 1.02^{-01}$ & $ 1.04^{-01}$ & $ 1.04^{-01}$ & $ 1.03^{-01}$\\
   7 &   11 & $ 6.77^{-02}$ & $ 6.93^{-02}$ & $ 7.18^{-02}$ & $ 7.45^{-02}$ & $ 7.70^{-02}$ & $ 7.88^{-02}$ & $ 7.96^{-02}$ & $ 7.94^{-02}$ & $ 7.81^{-02}$\\
   8 &   11 & $ 1.98^{-01}$ & $ 1.90^{-01}$ & $ 1.85^{-01}$ & $ 1.81^{-01}$ & $ 1.79^{-01}$ & $ 1.78^{-01}$ & $ 1.78^{-01}$ & $ 1.79^{-01}$ & $ 1.79^{-01}$\\   
   9 &   11 & $ 2.06^{-01}$ & $ 2.01^{-01}$ & $ 1.99^{-01}$ & $ 1.99^{-01}$ & $ 2.00^{-01}$ & $ 2.03^{-01}$ & $ 2.07^{-01}$ & $ 2.11^{-01}$ & $ 2.14^{-01}$\\
  10 &   11 & $ 3.32^{-01}$ & $ 3.24^{-01}$ & $ 3.19^{-01}$ & $ 3.17^{-01}$ & $ 3.17^{-01}$ & $ 3.17^{-01}$ & $ 3.16^{-01}$ & $ 3.14^{-01}$ & $ 3.09^{-01}$\\
 \bottomrule
 \end{tabular}
 \caption{Effective collision strengths as defined by equation (\ref{eq:ups}) are presented between an upper ($j$) and lower state ($i$), across a range of 9 temperatures (K), 3,800K $ < T < $ 25,100K.  \label{tab:ecollstrengths}}
\end{table*}

\end{document}